\documentclass[useAMS,usenatbib]{mn2e}

\usepackage{graphicx}
\usepackage{amssymb}
\usepackage{multirow}
\usepackage{pifont}
\usepackage{mathrsfs}

\makeatletter
    
    \newcommand{\Rmnum}[1]{\expandafter\@slowromancap\romannumeral #1@}
\makeatother

\title[Machine learning prediction for mean motion resonance behaviour -- The planar case] 
         {Machine learning prediction for mean motion resonance behaviour --  The planar case}
\author[Xin Li, Jian Li, Zhihong Jeff Xia and Nikolaos Georgakarakos]
{Xin Li$^{1}$, Jian Li$^{2,3}$\thanks{E-mail: ljian@nju.edu.cn}, Zhihong Jeff Xia$^{4}$ and Nikolaos Georgakarakos$^{5,6}$\\
$^1$ Department of Statistics and Data Science, Southern University of Science and Technology of China, No 1088, Xueyuan Rd., Xili, \\~~~Nanshan District, Shenzhen, Guangdong, 518055, PR China,\\
$^2$School of Astronomy and Space Science, Nanjing University, 163 Xianlin Avenue, Nanjing 210023, PR China,\\
$^3$Key Laboratory of Modern Astronomy and Astrophysics in Ministry of Education, Nanjing University, Nanjing 210023, PR China\\
$^4$Department of Mathematics, Northwestern University, 2033 Sheridan Road, Evanston, IL  60208, USA\\
$^5$New York University Abu Dhabi, PO Box 129188 Abu Dhabi, United Arab Emirates\\
$^6$Center for Astro, Particle and Planetary Physics (CAP$^3$), New York University Abu Dhabi, PO Box 129188 Abu Dhabi, United Arab Emirates}
\begin{document}

\date{Accepted 1988 December 15. Received 1988 December 14; in original form 1988 October 11}

\pagerange{\pageref{firstpage}--\pageref{lastpage}} \pubyear{2002}

\maketitle

\label{firstpage}

\begin{abstract}

Most recently, machine learning has been used to study the dynamics of integrable Hamiltonian  systems and the chaotic 3-body problem. In this work, we consider an intermediate case of regular motion in a non-integrable system: the behaviour of objects in the 2:3 mean motion resonance with Neptune. We show that, given initial data from a short 6250 yr numerical integration, the best-trained artificial neural network (ANN) can predict the trajectories of the 2:3 resonators over the subsequent 18750 yr evolution, covering a full libration cycle over the combined time period. By comparing our ANN's prediction of the resonant angle to the outcome of numerical integrations, the former can predict the resonant angle with an accuracy as small as of a few degrees only, while it has the advantage of considerably saving computational time. More specifically, the trained ANN can effectively measure the resonant amplitudes of the 2:3 resonators, and thus provides a fast approach that can identify the resonant candidates. This may be helpful in classifying a huge population of KBOs to be discovered in future surveys.

 \end{abstract}


\section{Introduction}

Neural networks were designed for pattern recognition by approximating the proper function between covariates (i.e. inputs) and variates (i.e. outputs) as a black-box \citep{McCu1943, Rosen1958}. Artificial neural networks (ANNs) were inspired by the human brain, but gradually became different from their biological prototype. This is quite similar to airplane design inspired by birds, but the principle was not taken from birds. Such a strategy has been efficiently applied to a wide range of pattern recognition problems in science and the industry. 
Usually the ANNs tackle large and complex machine learning tasks, such as classifying high quality pictures (e.g. ImageNet), recognising voice or speech (e.g. Siri), recommending goods to customers (e.g. YouTube, Amazon). Actually, the ANNs are data driven models and could possess a powerful ability to auto-explore the relation between inputs and outputs from the unknown physical systems. 

In machine learning, the convolutional neural networks (CNNs) are a class of the ANNs, most commonly applied to analyse the visual imagery \citep{valu20}. Because the computational power has increased significantly over the last few years, the CNNs have been managed to achieve superhuman performance in many complex visual tasks, such as for example in image search services \citep{bell2015,qay2017}, self-driving cars \citep{sanil20,stroe19} and automatic video classification \citep{karpathy2014,Ng2015}. More interestingly, the CNNs are also found to be successful at forecasting time series \citep{sezer20}. For a detailed description of CNNs, the reader is referred to the two books written by \citet{Geron2017} and \citet{Brownlee18}.

In the field of astronomy and astrophysics, the method of machine learning has already been widely applied to exoplanet discovery \citep{misl2018, scha19, arms21}, gamma-ray burst detection \citep{ukwa16, abra21}, galaxy classification \citep{Sun2019, zhan19, baqu21, vavi21}, the study of dark matter \citep{agar18, luci19, petu21}. For the classical n-body problem in celestial mechanics, however, the relative study did not start until very recently. \citet{Breen2020} intended to solve the general 3-body problem by means of training a \textit{deep ANN} (i.e. with more than 1 hidden layer). For the planar case, they demonstrate that the deep ANN can provide solutions as accurate as those from the numerical integration, but cost much less computational time. However, we notice that their best-performing ANN is trained for trajectories with an evolution time of $T=3.9$ time units. By adopting  similar dimensionless units in a simplified Solar system that consists of the Sun, a planet and an asteroid, this time interval is merely of the order of hundreds of years, while for longer timescales of $T=7.8$ or $T=10$, the loss on the validation set would become much larger.

The strong dependence of the accuracy of the ANN's solution on the evolution time of the 3-body system is quite straightforward because this dynamical system is chaotic \citep{Poin1892}. Chaotic systems are extremely sensitive to the initial conditions, as slight changes in the input can lead to significantly different long-term output. Thus, it is essentially difficulty to predict the trajectory evolution longer than the Lyapunov time, i.e. the inverse of the Lyapunov characteristic exponent. On the other hand, as the prediction accuracy is also sensitive to the patterns of the training and validation data sets \citep{Breen2020}, the construction of the ANNs for the chaotic problem is therefore a complicated task. \citet{Grey2019} studied integrable Hamiltonian systems where the motion is regular everywhere, e.g. the ideal mass-spring and the ideal pendulum. For such systems, they designed the so-called Hamiltonian neural networks to learn the laws of physics directly from data, and also the conservation of energy-like quantities on a long timescale.

Moreover, there exists some literature on machine learning in orbital dynamics. \citet{Tamayo2016} showed that characterising the complicated and multi-dimensional stability boundary of tightly packed systems is amenable to an XGBoost machine learning algorithm. \citet{Tamayo2020} combined analytical understanding of resonant dynamics in two-planet systems with machine-learning techniques to train a model capable of robustly classifying stability in compact multi-planet systems over long timescales of $10^9$ orbits. \citet{Cranmer2021} introduced a Bayesian neural network model to predict when a compact planetary system with three or more planets will go unstable. Their model was trained directly from short-time series of raw orbital elements, as well as from  works related to orbital stability \citep{Lam2018, bha21}. We would like to point out that, together with \citet{Breen2020}, all those similar works only study the planar case. This choice is quite reasonable because it requires lower computational capacity than in the spacial case.

\subsection{The Kuiper belt and machine learning study}

In the outer Solar system beyond the orbit of Neptune, there is a large number of icy celestial bodies, known as the Kuiper belt objects (KBOs). A striking feature is that the population in mean motion resonance (MMR) with Neptune comprises a high proportion of the observed KBOs \citep{Glad2008, Khain20}. By now, about 400 objects are found to accumulate in the 2:3 MMR around 39.4 au, and have eccentricities up to 0.3. Their dynamical properties are important clues to the early history of our planetary system. For instance, these objects provide solid evidence that the Jovian planets experienced substantial orbital migration \citep{malh93, malh95}. While migrating outwards, Neptune had its 2:3 resonance capturing a lot of planetesimals including Pluto, and also exciting their eccentricities. Since the 1990s, the origin and evolution of the resonant KBOs have received considerable attention \citep{malh05, li11, nesy16, volk16, pike17, yu18, lawl19}. As a result of the ongoing surveys, much more KBOs are expected to be discovered in the near future, e.g. the Large Synoptic Survey Telescope (LSST) will make a significant contribution by discovering about 30,000 KBOs \citep{ivez19}. Then the refined structure of the resonant population would help to further improve the planet migration scenario.

\begin{figure}
 \hspace{0cm}
  \centering
  \includegraphics[width=9cm]{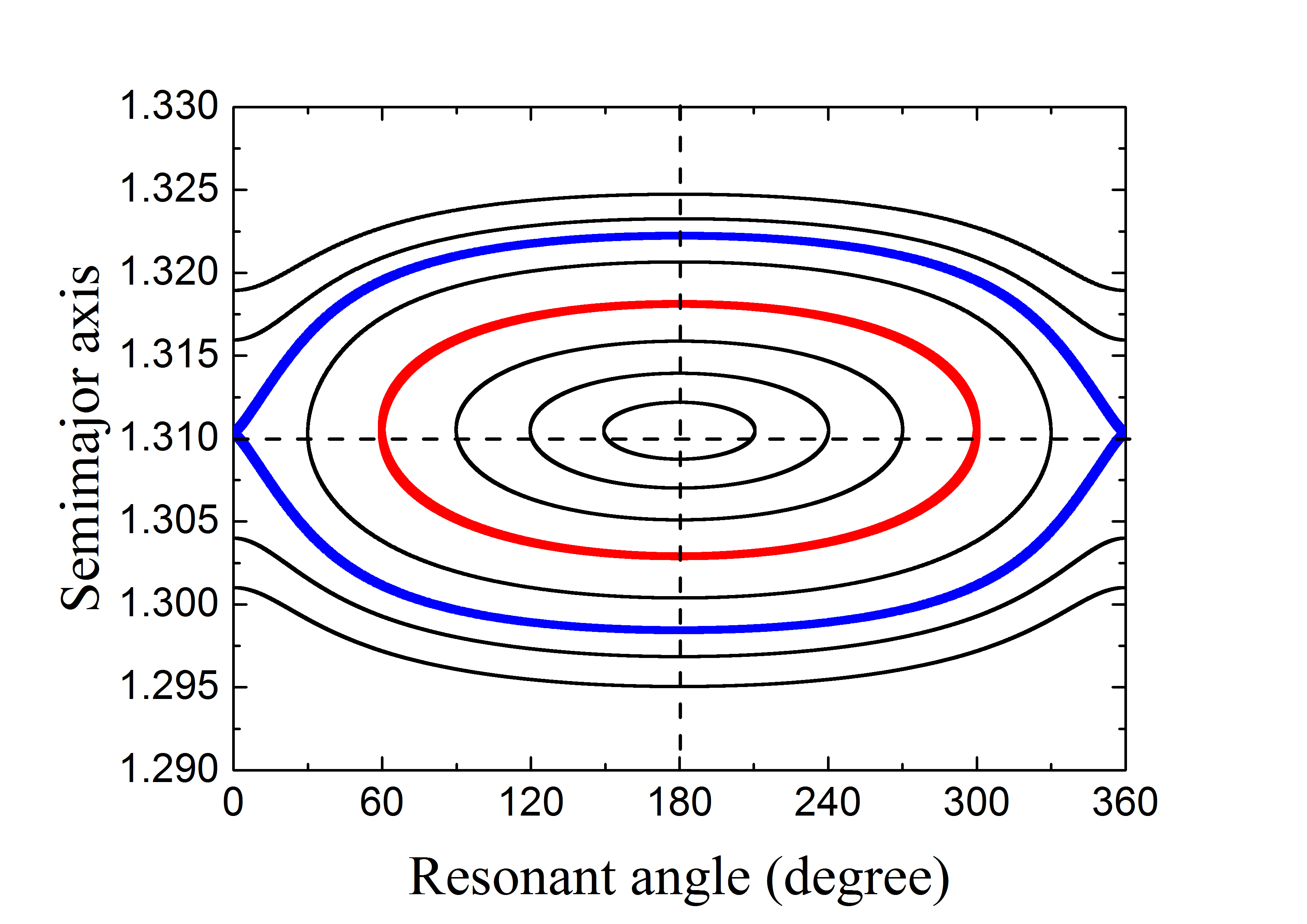}
\caption{The phase space of Neptune's 2:3 MMR for particle's eccentricity $e=0.1$ in the framework of the planar circular restricted 3-body problem. In this panel, the semimajor axis is scaled by Neptune's semi-major axis. The blue curve indicates the separatrix between libration and circulation, and the red curve in the libration island denotes the contour of the resonant amplitude of $120^{\circ}$. For reference, the horizontal dashed line is plotted at the location of the nominal 2:3 MMR, and the vertical dashed line depicts the libration centre at $180^{\circ}$.}
  \label{pendu}
\end{figure}

 In the framework of the planar circular restricted 3-body problem (PCR3BP), i.e. Sun+Neptune+particle, we intend to explore the performance of the ANN on learning the dynamical evolution of the particles in the 2:3 resonance with Neptune. As for adopting the planar model, the computational limitation is one consideration as we noticed in the aforementioned works; besides, this 2:3 resonance is actually an eccentric-type resonance \citep{Li2020}, whose dynamical features are almost determined by the particle's eccentricity, but not the inclination. Although the PCR3BP is the simplest model in the 3-body problem, it contains both chaotic and regular motions. When the eccentricity of the particle is not too large, there are only invariant tori existing in the phase space of the 2:3 MMR (see Fig. \ref{pendu}), and thus the motion is always regular. The first goal of this work is to train an ANN, over a long and fixed time interval, to accurately predict the behaviour of the regular resonant motion.

Our second goal is to provide a fast identification approach to the resonant KBOs. As the trained ANN may predict the evolution of the particles considerably faster than numerical integrations, we can  measure the resonant amplitudes accordingly within quite a short computing time. The machine learning classification of resonant populations in the Solar system did not start until last year. \citet{smul20} used a decision tree classifier trained on various orbital features to sort the KBOs into four dynamical classes: the classical, resonant, detached and scattering objects. In addition, regarding the main belt asteroids in the 1:2 MMR with Mars, \citet{Carruba2021} applied ANN to recognise the images of the time-varying resonant angles, then to classify the libration, switching and circulating orbits. In the above two works, a $10^5$ yr numerical integration is required for generating the trajectories to be identified. We aim to further reduce the time interval of the numerical integration and let the ANN to make the prediction of the considered trajectories over a timescale as long as possible.

The rest of this paper is organised as follows: in Section 2, we briefly introduce the PCR3BP model, and how to generate the samples in Neptune's 2:3 MMR for the training and validation sets. In Section 3, we describe how to design and train our ANNs, and show the dependence of the results on the structure of the training set. In Section 4, we apply our trained ANNs to particles in the stable 2:3 resonance, for predicting orbital evolution and identifying resonant behaviour. Finally, the conclusions and discussion are given in Section 5 and Section 6, respectively.

\section[]{Learning data generation for the 2:3 resonance}

\subsection{The PCR3BP}

We adopt the PCR3BP in order to model the particles in the 2:3 MMR with Neptune and generate the data sets for machine learning. In the framework of this dynamical model, we refer to the Sun with mass $m_1$ as the primary, the planet Neptune with mass $m_2$ as the secondary, and a massless particle as the third body. The primary and secondary exert gravitational forces on the particle but they are not affected by the particle; the motions of all three bodies take place in the same plane.

Since the Sun and Neptune are moving on circular orbits about their common centre of mass $O$, they have a constant mutual distance $a_N$ and the same angular velocity $n_N$. It is customary to choose the non-dimensional parameters in the following way: the total mass of the system $(m_1 + m_2)=1$, the separation $a_N=1$, and the angular velocity $n_N=1$. Given these units, the constant of gravitation is also equal to 1. We employed the synodic coordinate system $(x, y)$, which has the origin located at $O$ but is rotating at a uniform rate $n_N$ in the positive direction of Neptune's orbital motion. The $x$-axis is chosen such that the Sun and Neptune always lie along it with coordinates $(-\mu, 0)$ and $(1-\mu, 0)$, respectively, where the mass ratio $\mu=m_2/(m_1 + m_2)\approx5.146\times10^{-5}$.

In the synodic system $(x, y)$, the equations of the motion of the massless particle are 
\begin{eqnarray}
\ddot{x}-2\dot{y}&=&\frac{\partial U}{\partial x},\nonumber\\
\ddot{y}+2\dot{x}&=&\frac{\partial U}{\partial y},
\label{3b}
\end{eqnarray}
where the ``pseudo-potential'' $U=U(x, y)$ is given by
\begin{equation}
U=\frac{1}{2}(x^2+y^2)+\frac{1-\mu}{r_1}+\frac{\mu}{r_2},
\label{Ufunction}
\end{equation}
and $r_1$ and $r_2$ are the particle's distances to the Sun and Neptune, respectively:
\begin{eqnarray}
r_1^2&=&(x+\mu)^2+y^2,\nonumber\\
r_2^2&=&(x+\mu-1)^2+y^2.
\label{distance}
\end{eqnarray}

The differential system given in equation (\ref{3b}) has a constant of motion, known as the Jacobi integral: 
\begin{equation}
C(x, y, \dot{x}, \dot{y})=2U(x, y)-(\dot{x}^2+\dot{y}^2),
\label{jacobi}
\end{equation}
which will be used to place restrictions on the motion of the particles for either the numerical integration or the machine learning prediction.

\subsection{The 2:3 resonant samples}

We regard a particle locked in Neptune's 2:3 MMR if the critical resonant argument
\begin{equation}
\sigma=3\lambda-2\lambda_N-\varpi,
\label{ResAng}
\end{equation}
librates around $180^{\circ}$, where $\lambda$ and $\varpi$ are the mean longitude and the longitude of perihelion of the particle, and $\lambda_N$ is the mean longitude of Neptune.

For the sake of generating a resonant population in Neptune's 2:3 MMR, the initial conditions of the particles are given in terms of the orbital elements $(a, e, \varpi, \lambda)$, where $a$ is the semimajor axis and $e$ is the eccentricity. We sampled particles starting at the location of the nominal 2:3 resonance, defined as $a_{res}=\sqrt[3]{3^2/2^2}\cdot a_N\approx1.31$ \citep{Gall2006}. Note that in the PCR3BP, the resonant angle $\sigma$ is invariant to changes in $\varpi$ (denoted by $\Delta \varpi$) if we simply rotate the coordinate system about the origin $O$ by the angle $\Delta \varpi$, so $\varpi$ is taken to be an arbitrary value of $60^{\circ}$. Then the initial value of $\lambda$ can be determined from a given $\sigma$ by equation (\ref{ResAng}).

As shown in Fig. \ref{pendu}, at a relatively small $e=0.1$, the phase space of the 2:3 MMR is regular everywhere, and there is only a symmetric libration centre having $a_{res}=1.31$ and $\sigma=180^{\circ}$. Since $a$ is chosen to be $a_{res}$, the initial resonant angle $\sigma_0$ measures the resonant amplitude $A_{\sigma}$, i.e. the maximum deviation of $\sigma$ from the resonant centre at $180^{\circ}$. It is important to mention that a large fraction of the observed 2:3 resonant KBOs are on more eccentric orbits (e.g. Pluto with $e\sim0.25$), for which the chaotic motion may appear at large $A_{\sigma}$ \citep{malh96}. Nevertheless, according to our previous work \citep{Li2014}, the 2:3 resonant KBOs with $e$ up to 0.3 could be stable over the age of the Solar system if their resonant amplitudes $A_{\sigma} < 120^{\circ}$. Thus, for the resonant samples introduced to generate the machine learning data, their initial resonant angles $\sigma_0$ will be restricted to the region of $[180^{\circ}-120^{\circ}, 180^{\circ}+120^{\circ}]$, which is enclosed by the red contour of $A_{\sigma}=120^{\circ}$ in Fig. \ref{pendu}. For the remainder of this paper, first, we consider the case of $e=0.1$ for the 2:3 resonators. Provided the ANN is effective for predicting their resonant behaviours, we will generalise our results to the more eccentric cases of $e=0.2$ and $e=0.3$.

For a sampled particle, when a set of initial $(a, e, \varpi, \lambda)$ is given, we can calculate the corresponding initial positions $(x, y)$ and velocities $(v_x, v_y)$. Then we numerically propagate the equations of motion, i.e. equations (\ref{3b}), in the synodic frame over a timescale of $T_{tot}=25000$ yrs. The numerical integration is performed by using the 8th-order Runge-Kutta integrator with a time step of $\sim0.01$ yr, which can keep a desired degree of error less than $10^{-20}$. The adopted timescale $T_{tot}$ is equivalent to 100 orbits at the location of Pluto, and it is long enough to see a full libration cycle for the typical 2:3 resonant particles. Therefore, we can determine the particle's resonant amplitude $A_{\sigma}$, as well as the resonant or non-resonant behaviour. The time interval between successive outputs in the integration is chosen to be $T_{tot}$/99, such that a trajectory comprises a data set of 100 discrete points ($x(t_i),y(t_i),v_x(t_i),v_y(t_i)$) at the respective time $t_i$ ($i=1,...,100$). Accordingly, we can deduce the orbital elements ($a(t_i),e(t_i),\varpi(t_i),\lambda(t_i)$), which are in the heliocentric inertial frame.

Through the above process, we generate a training set containing 10000 trajectories whose initial resonant angles $\sigma_0$ are randomly chosen between $60^{\circ}$ and $300^{\circ}$, i.e. $A_{\sigma} \le 120^{\circ}$. For each trajectory, there are 100 orbital points characterised by four covariants $(a(t_i),e(t_i),\varpi(t_i),\lambda(t_i))_{i=1,...,100}$ with the same time interval. This first experiment will be referred as \textit{Train} \Rmnum1 (see Table \ref{trainset}). While for the validation set, we use 1000 trajectories starting within the same $\sigma_0$ region, given a uniform distribution with a resolution of $\Delta\sigma_0=0.24^{\circ}$. In this way, the trajectories in the validation set could be close to those in the training set, but would not overlap.


\section{Learning method}

\begin{figure}
 \hspace{0cm}
  \centering
  \includegraphics[width=8cm]{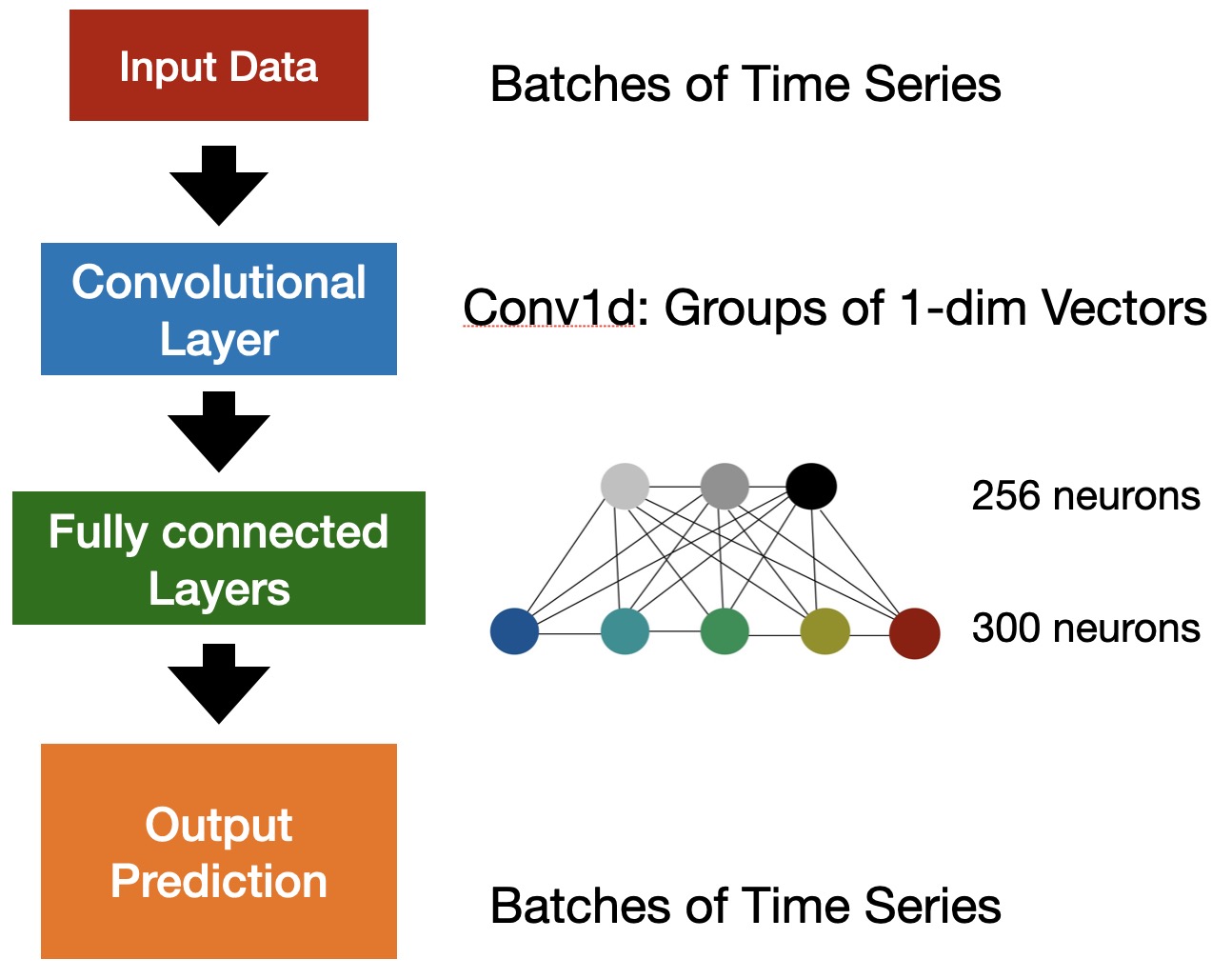}
  \caption{ We constructed an ANN composed of the input layer, one convolutional layer, two fully connected layers and the output layer. The actual numbers of the neurons on individual fully connected layers are depicted on the side. We set initial weights on all the connected links. After every training epoch, the weights are renewed, then the refined parameters of the ANN model will be automatically saved. }
  \label{ann}
\end{figure}

\subsection{The ANN initialization}

Aiming to predict the resonant motions of the particles in the PCR3BP, we construct the ANN to learn from the training set (i.e. \textit{Train} \Rmnum1), and then evaluate the prediction accuracy by using the validation set. The structures of these two sets have been described just above. In order to test whether the ANN can make accurate predictions for the resonant behaviours, we focus on the particles with small $e=0.1$ in the first place. As displayed in Fig. \ref{pendu}, these particle trajectories are restricted on the invariant tori in the libration zone and therefore their motions are considered to be regular.

All the following experiments are taken in Pytorch. We constructed a feed-forward ANN which consists of 5 layers: 1 input layer, 1 convolutional layer, 2 fully connected layers and 1 output layer (Fig. \ref{ann}). The convolutional layer and the fully connected layers are usually called \textit{hidden layers}. The convolutional layer is the most important building block of a CNN, as we mentioned in the introduction. Within the Pytorch framework, we applied the 1-dimensional convolutional layer (Conv1d), set the size of input and output channels to be 4 and 10 respectively, and took the kernel size of convolution to be 1. The quantity of input channels is equal to the dimension of the multivariate set $(a,e,\varpi,\lambda)$ and the amount of output channels is the number of convolutions in the layer. The next two fully connected layers have 256 and 300 neurons respectively. In all the tasks, we trained the models with a learning rate of $10^{-4}$ and adopted the Adam optimizer \citep{Ba2015}.

The weights are initially assigned from a Gaussian distribution on all hidden layers of the network. Considering the computational capacity, we set the batch size to be 100, i.e. we separate the training data into 100 batches. This trick could also be helpful to efficiently reduce the loss during the learning process. We choose the usual Rectified Linear Unit (ReLU) activation function $\psi(z)=max(0, z)$ at all hidden layers \citep{Hinton2010}.  Suppose that we have input data $x_1,x_2,x_3,...,x_n$, and the connection weights between the input and hidden layers are $w_1,w_2,w_3,...,w_n$. Then the intermediate variable is $z=\sum_{i=1}^nw_i\cdot x_i$,  and we have the activation function $\psi(z)$ acting on $z$. The activation function includes the nonlinear effects of the problem to be solved, thus it is a key point for ANN to converge easier. We also tried different activation functions, such as the leaky rectified linear unit function \citep{Maas2013} and the exponential rectified function \citep{Clevert2016}. However, no apparent improvement was achieved.

\subsection{Training process}

Next, the network training has been carried out, by learning from the training data and tuning-up the weights to reduce the loss. Each training epoch contains a complete forward-reverse pass over the training data. This training strategy is well known as \textit{Backpropagation} \citep{Rume1986}.

The forward pass performs at first. As for the 100 time-points of each trajectory, we enter the first $m$ points ($a(t_i),e(t_i),\varpi(t_i),\lambda(t_i)$) ($1\le i\le m$) into the input layer. The algorithm computes the intermediate result of every neuron in the hidden layers. Finally the output layer returns the following $n=100-m$ predicted points $(\hat a(t_i),\hat e(t_i),\hat \varpi(t_i),\hat \lambda(t_i))$ ($m+1 \le i \le 100$). The accuracy of the ANN can be measured by the loss $\mathcal{L}$. By comparing with the labels $(a(t_i),e(t_i),\varpi(t_i),\lambda(t_i))$ at the corresponding time $t_i$ ($m+1 \le i \le 100$), the loss is defined as
\begin{eqnarray}
\mathcal{L}& =\frac{1}{4*(100-m)}\sum_{i=m+1}^{100}\left(|\hat a(t_i)-a(t_i)|+|\hat e(t_i)-e(t_i)|\right. \nonumber\\
        & \left. +|\hat \varpi(t_i)-\varpi(t_i)|+|\hat \lambda(t_i)-\lambda(t_i)|\right).
\label{L1loss}
\end{eqnarray}
Hereafter, the losses on the training and validation sets would be denoted by $\mathcal{L}_{train}$ and $\mathcal{L}_{valid}$, respectively.

Then the reverse pass starts. Given each neuron's loss at the output layer, our algorithm computes backwards the contribution $B^{(1)}_j$ of each neuron at the final hidden layer, through the connections between these two layers by the chain rule. Then, regarding the obtained $B^{(1)}_j$, the algorithm continues to evaluate the contribution $B^{(2)}_j$ for each neuron at the previous hidden layer. This process stops when the calculation of such contribution reaches the input layer. By now, we get the loss gradient backwards across all the weights on the connection links. At the stage of the reverse pass, the Adam optimization is adopted to lessen the loss by adjusting the connection weights.

\begin{figure}
 \hspace{0cm}
  \centering
  \includegraphics[width=9cm]{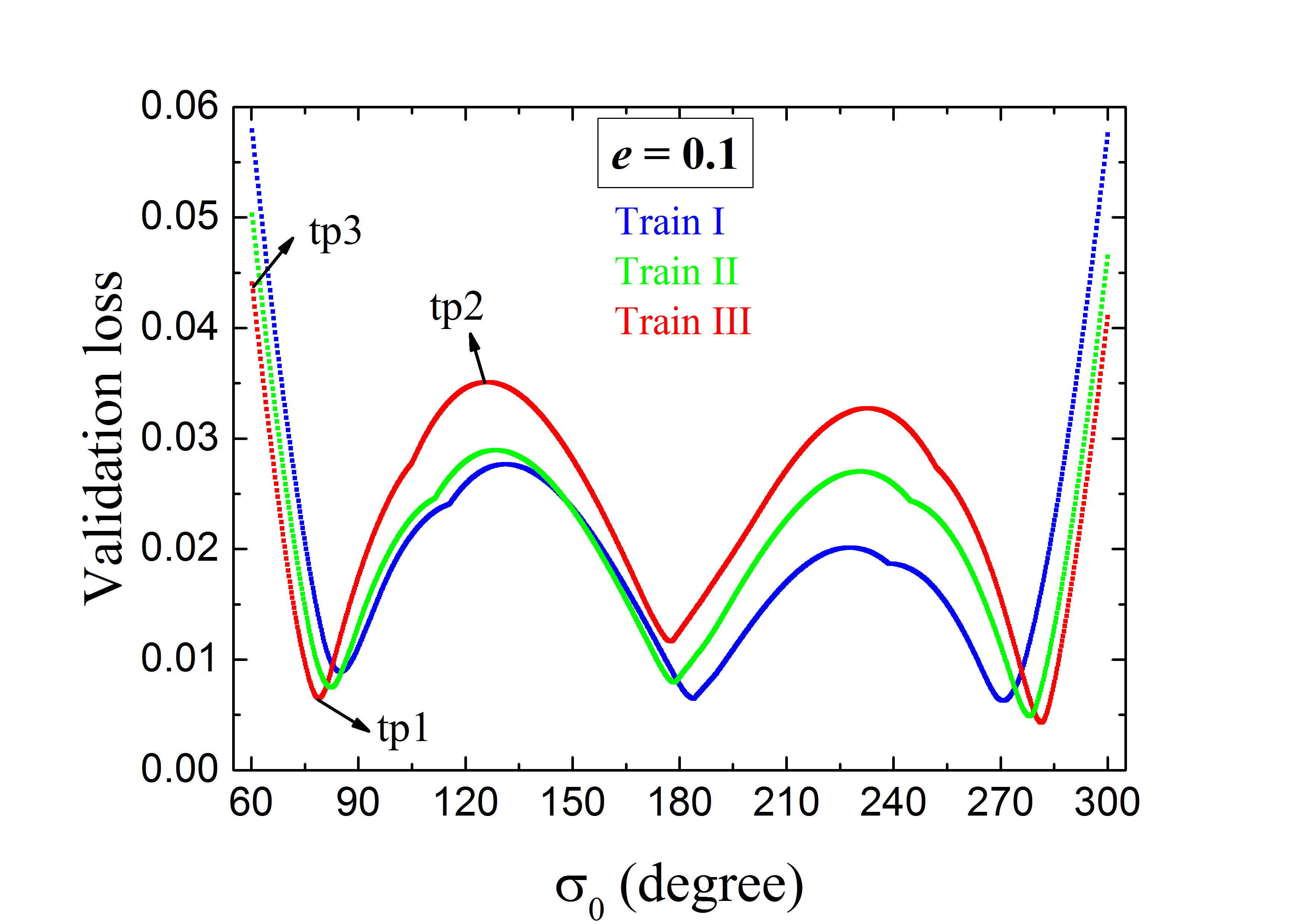}
  \caption{In the $e=0.1$ case, the loss on the validation set consists of samples with initial resonant angles $\sigma_0$ symmetrically distributed with respect to the resonant centre at $180^{\circ}$. The ANNs are trained for 3000 epochs, by using the training sets \textit{Train} \Rmnum1 (blue),  \textit{Train} \Rmnum2 (green) and \textit{Train} \Rmnum3 (red). For reference, in the case of \textit{Train} \Rmnum3, we indicate three typical examples which have the minimum loss (tp1 with $\sigma_0=79^{\circ}$), local maximum loss (tp2 with $\sigma_0=126^{\circ}$) and maximum loss (tp3 with $\sigma_0=60^{\circ}$), respectively.}
  \label{loss}
\end{figure}

\begin{table*}
\centering
\begin{minipage}{15cm}
\caption{The initial conditions of the resonant angle $\sigma_0$, the resonant amplitude $A_{\sigma}$, the number density $\Sigma$ as a function of $A_{\sigma}$ and the corresponding cumulative number for particles in three different training sets.}      
\label{trainset}
\begin{tabular}{l c c c c}        
\hline                 
Training set            &       $\sigma_0$           &    $A_{\sigma}(>0)$     &    $\Sigma$                      &     Cumulative number                                    \\

\hline

\textit{Train} \Rmnum1  &  [$-120^{\circ}$, $120^{\circ}$]    &  $\le120^{\circ}$   &  ~~~~~~~~~~~~~~~~~~~~$\propto (A_{\sigma})^{0}$ (i.e. constant)                   &    $\propto (A_{\sigma})^{1}$    \\

\textit{Train} \Rmnum2  &  [$-120^{\circ}$, $120^{\circ}$]    &  $\le120^{\circ}$   &   $\propto (A_{\sigma})^{1}$    &    $\propto (A_{\sigma})^{2}$                         \\

\textit{Train} \Rmnum3  &  [$-120^{\circ}$, $120^{\circ}$]    &  $\le120^{\circ}$   &  $\propto (A_{\sigma})^{2}$                        &    $\propto (A_{\sigma})^{3}$                      \\

\hline
\end{tabular}
\end{minipage}
\end{table*}

During the training process, one needs to refine all the hyperparameters (e.g. the choice of activation functions and the number of the hidden layers or neurons), to make sure the training loss $\mathcal{L}_{train}$ can decrease overall as the number of epochs increases, and the precision of prediction on the training set can reach a reasonable level.  As a result, the ReLU activation function was chosen, and the structure of the ANN showed in Fig. \ref{ann} was determined. Suppose the training is at epoch $k$, if the obtained loss $\mathcal{L}_{train}^k$ is smaller than the previous ones $\mathcal{L}_{train}^{k-1},\dots, \mathcal{L}_{train}^0$, all the connection weights of the ANN will be saved.

There is an important parameter $m$ to be determined. The value of $m$ scales with how fast a trajectory can be obtained. Our first thought is to try $m=1$, i.e. to use only the first time point of a trajectory to predict the following 99 ones. But in this case, we find that the loss $\mathcal{L}_{train}$ has ceased to decrease when it still has a relatively large value. It is easy to understand that because the input information is insufficient to characterise the evolution of the resonant motion. Thus, we have to choose a larger value of $m$. After some experiments, we decide to adopt $m=25$, which is about the length of the 1/4 of the libration period and results in a much more accurate prediction of the following 75 time points. In this circumstance, we can save the computational time at the level of $75/(25+75)=3/4$, i.e. 75\%. Actually, if we use an even larger value for $m$, such as $m=99$, $\mathcal{L}_{train}$ could certainly further decrease, but only one time point is to be predicted. In this case, the machine learning method can not effectively speed up the orbital computation, failing to meet the main goal of our work.

We also monitor the variation of the validation loss $\mathcal{L}_{valid}$ throughout our computation. As a matter of fact, at each epoch, the ANN automatically calculates both $\mathcal{L}_{train}$ and $\mathcal{L}_{valid}$. The result shows a general trend of a decrease of $\mathcal{L}_{valid}$ with increasing epoch during the training process. Here, we need to clarify that $\mathcal{L}_{valid}$ has never been used to make any decision on the structure of the ANN or the choices of all the connection weights, while it only represents the generalisation ability of the trained ANN. The profile of the loss $\mathcal{L}_{valid}$ will be presented in the following subsection.

\subsection{The ANN validation}

In all our tasks, after 3000 epochs, the training losses turn out to be quite small and have nearly ceased to decrease. Meanwhile the validation losses don't seem to increase, indicating that there is no issue of overfitting the data. Thus, we suppose that the  prediction accuracy could be acceptable, and all the data were  recorded at the very epoch 3000. Fig. \ref{loss} shows, for the ANN trained by the set \textit{Train} \Rmnum1 (blue curve), the validation loss $\mathcal{L}_{valid}$ as a function of the initial resonant angle $\sigma_0$. We can see that the values of $\mathcal{L}_{valid}$ are quite small ($\sim0.01$) near the resonant centre at $\sigma_0=180^{\circ}$. Although there are two peaks around $\sigma_0=120^{\circ}$ and $240^{\circ}$, their heights are just of the order of $\sim0.03$. However, $\mathcal{L}_{valid}$ could notably increase for the particles starting farthest away from the resonant centre, and the maximum $\mathcal{L}_{valid}$ is as large as $\sim0.06$ at both ends of the (blue) loss curve. Even we continue to run more epochs, there is little decline in the maximum $\mathcal{L}_{valid}$ around the two boundary points of the $\sigma_0$-interval. To overcome this drawback, we also made efforts to look for a better locally optimal solution for the ANN training, by adopting some other optimisation algorithms such as an adaptive gradient algorithm called AdaGrad \citep{Duchi2011} and the stochastic gradient descent (SGD) method. However, their performances are even not as good as the Adam optimizer we chose earlier.

We then realize that, the largest $\mathcal{L}_{valid}$ appearing farthest away from the resonant centre could be owing to insufficient input information for the corresponding resonant particles. In the training set \textit{Train} \Rmnum1, the samples have their initial resonant angles $\sigma_0$ evenly distributed with respect to the resonant centre at $180^{\circ}$. As we noted previously, the resonant amplitude $A_{\sigma}$ can simply be measured by the deviation of $\sigma_0$ from $180^{\circ}$. It means that the number density $\Sigma$ is proportional to $(A_{\sigma})^{0}$, i.e. a constant. We then expect that this issue could be fixed by increasing the number density $\Sigma$ at larger $A_{\sigma}$. Consequently, inside the considered resonant zone, we construct two other training sets, which consist of particles with $\Sigma$ increasing as $(A_{\sigma})^{1}$ and $(A_{\sigma})^{2}$. These two sets would be referred as \textit{Train} \Rmnum2 and \textit{Train} \Rmnum3 (see Table \ref{trainset}). For \textit{Train} \Rmnum2 and \textit{Train} \Rmnum3, the cumulative numbers of particles within a certain torus characterized by $A_{\sigma}$, are  respectively proportional to $(A_{\sigma})^{2}$ and $(A_{\sigma})^{3}$. Besides the training data, the ANN's structure and the validation set are kept the same and the results will be presented in the next section.


\section{Results}

\subsection{The case of $e=0.1$}

\begin{figure*}
  \centering
  \begin{minipage}[c]{1\textwidth}
  \vspace{0 cm}
  \includegraphics[width=9cm]{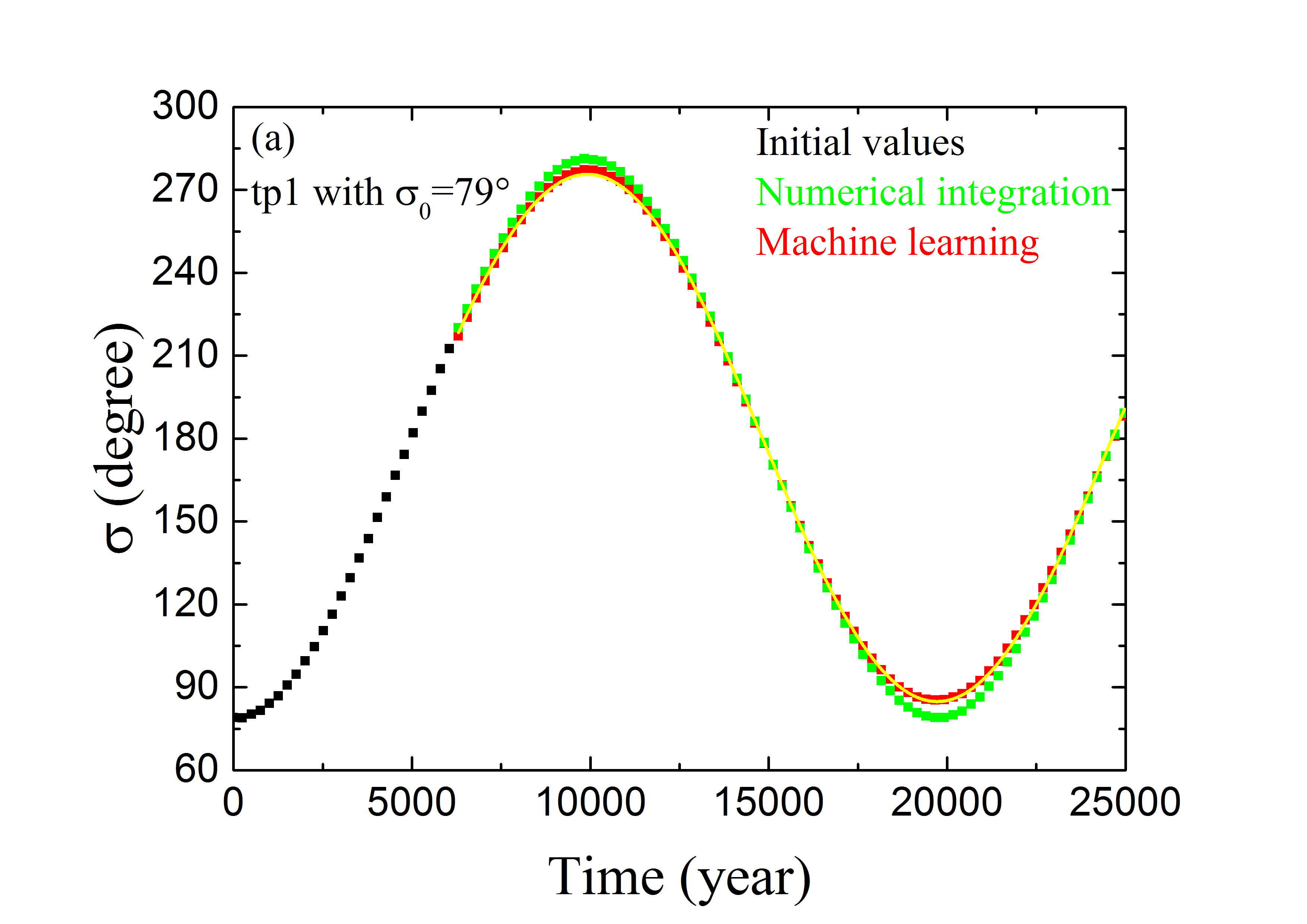}
  \includegraphics[width=9cm]{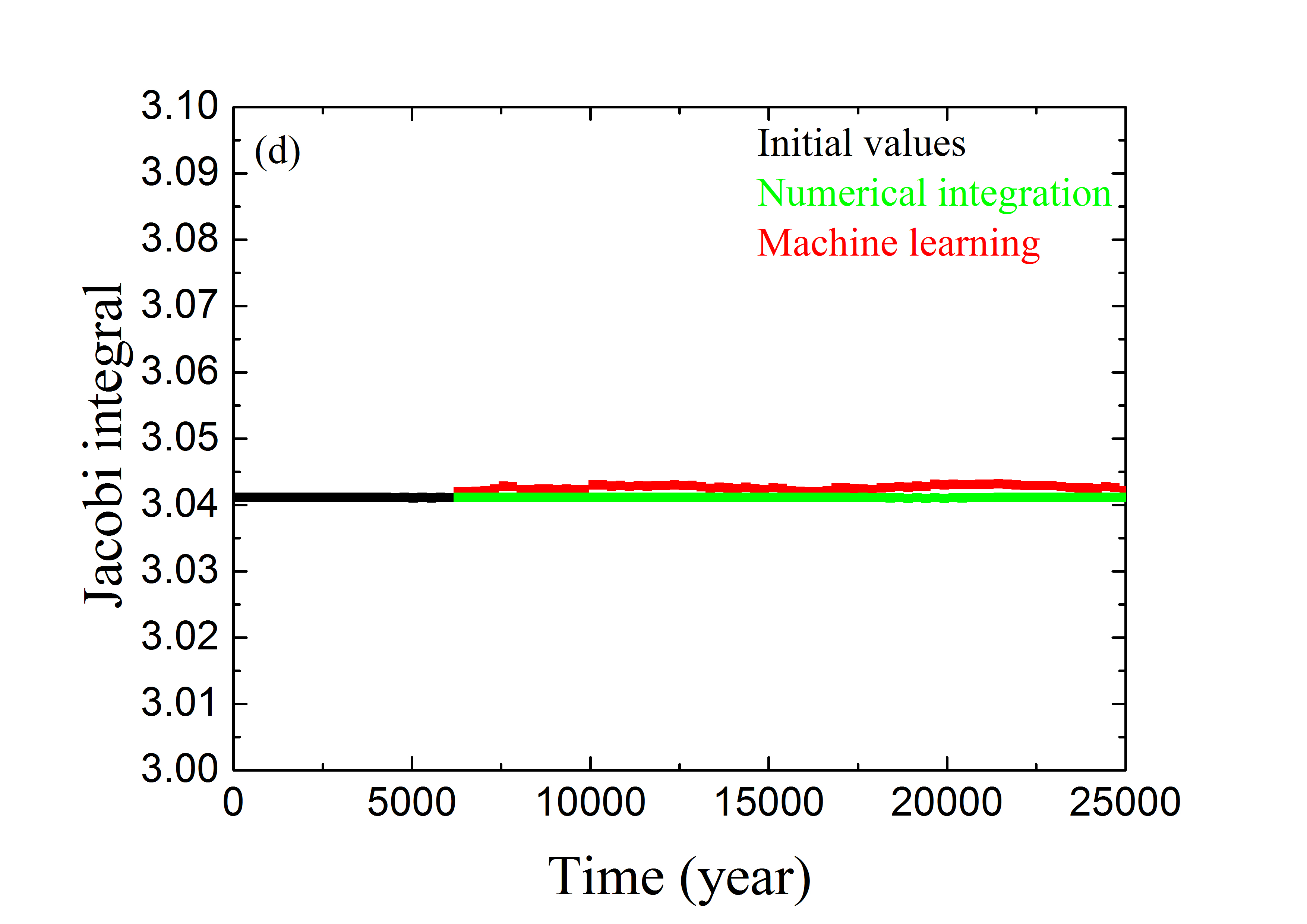}
  \end{minipage}
  \begin{minipage}[c]{1\textwidth}
  \vspace{0 cm}
  \includegraphics[width=9cm]{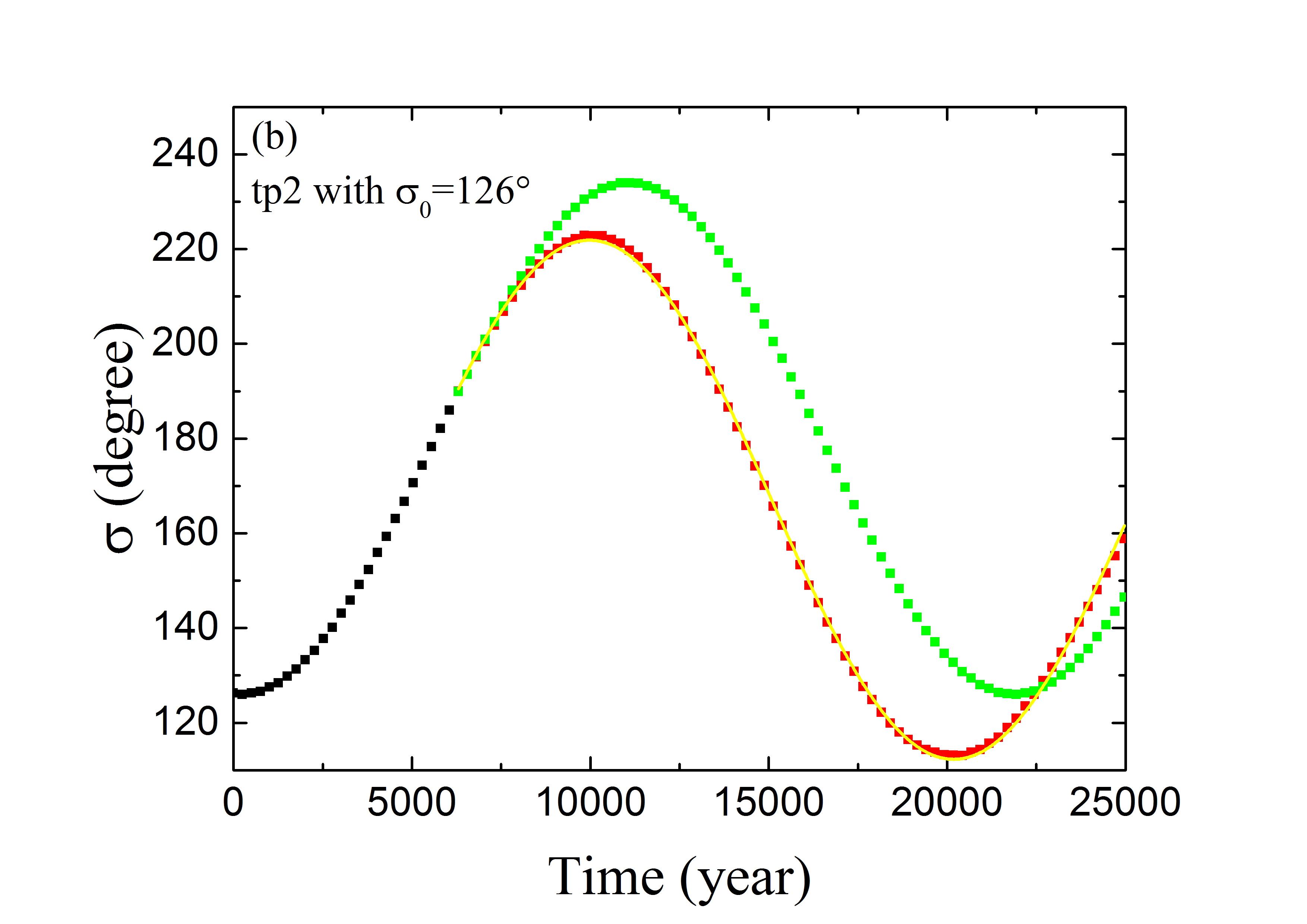}
  \includegraphics[width=9cm]{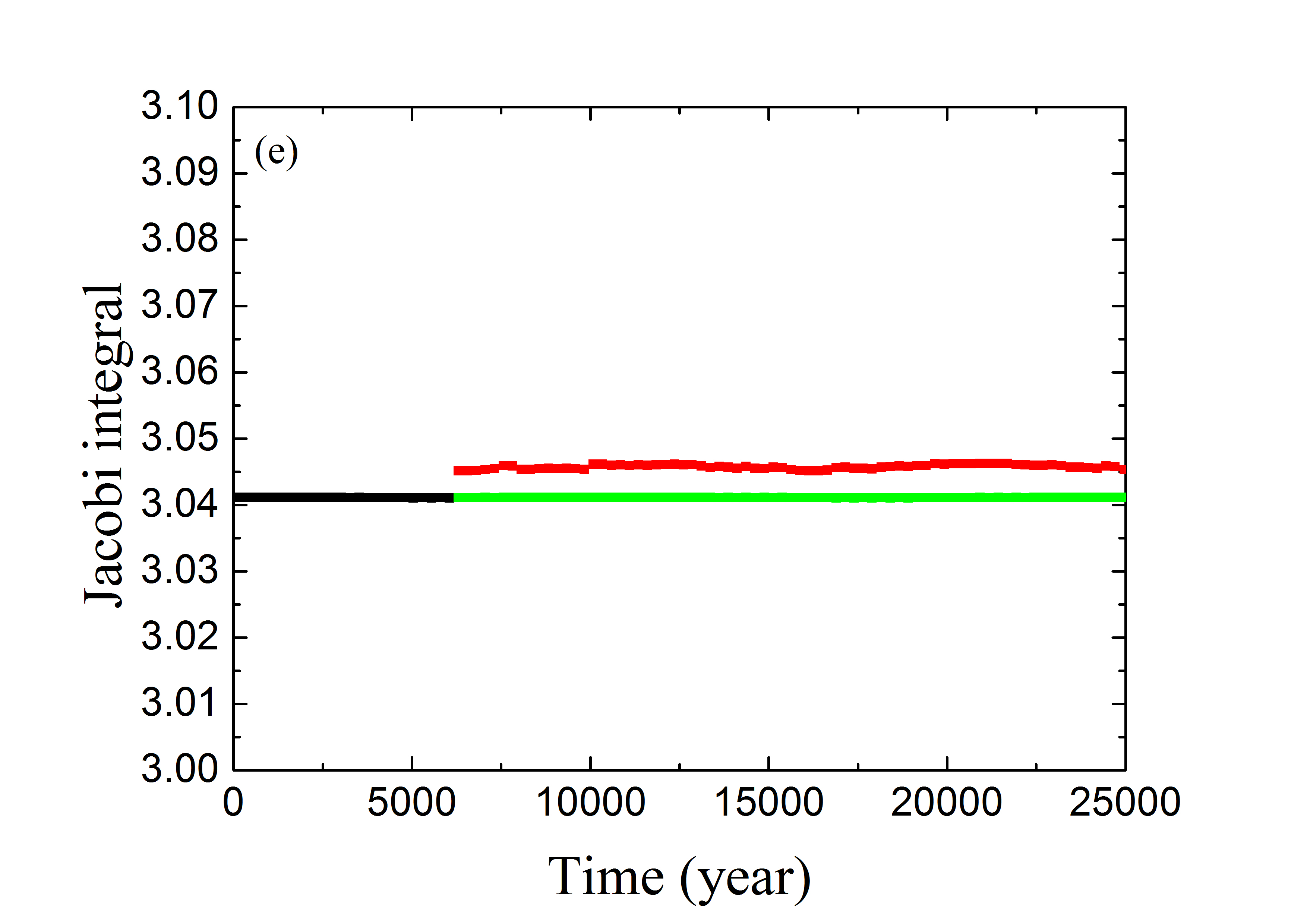}
  \end{minipage}
  \begin{minipage}[c]{1\textwidth}
  \vspace{0 cm}
  \includegraphics[width=9cm]{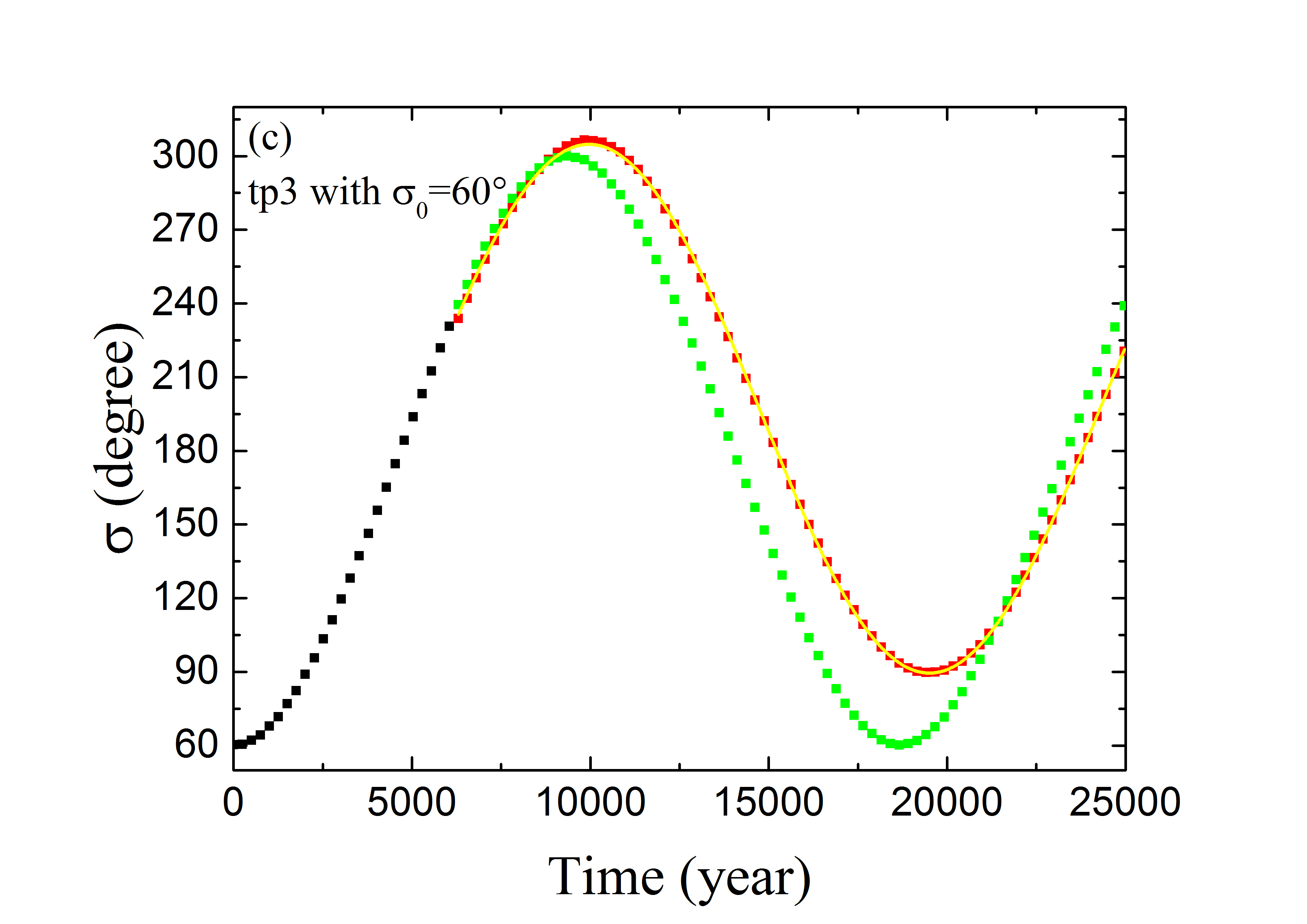}
  \includegraphics[width=9cm]{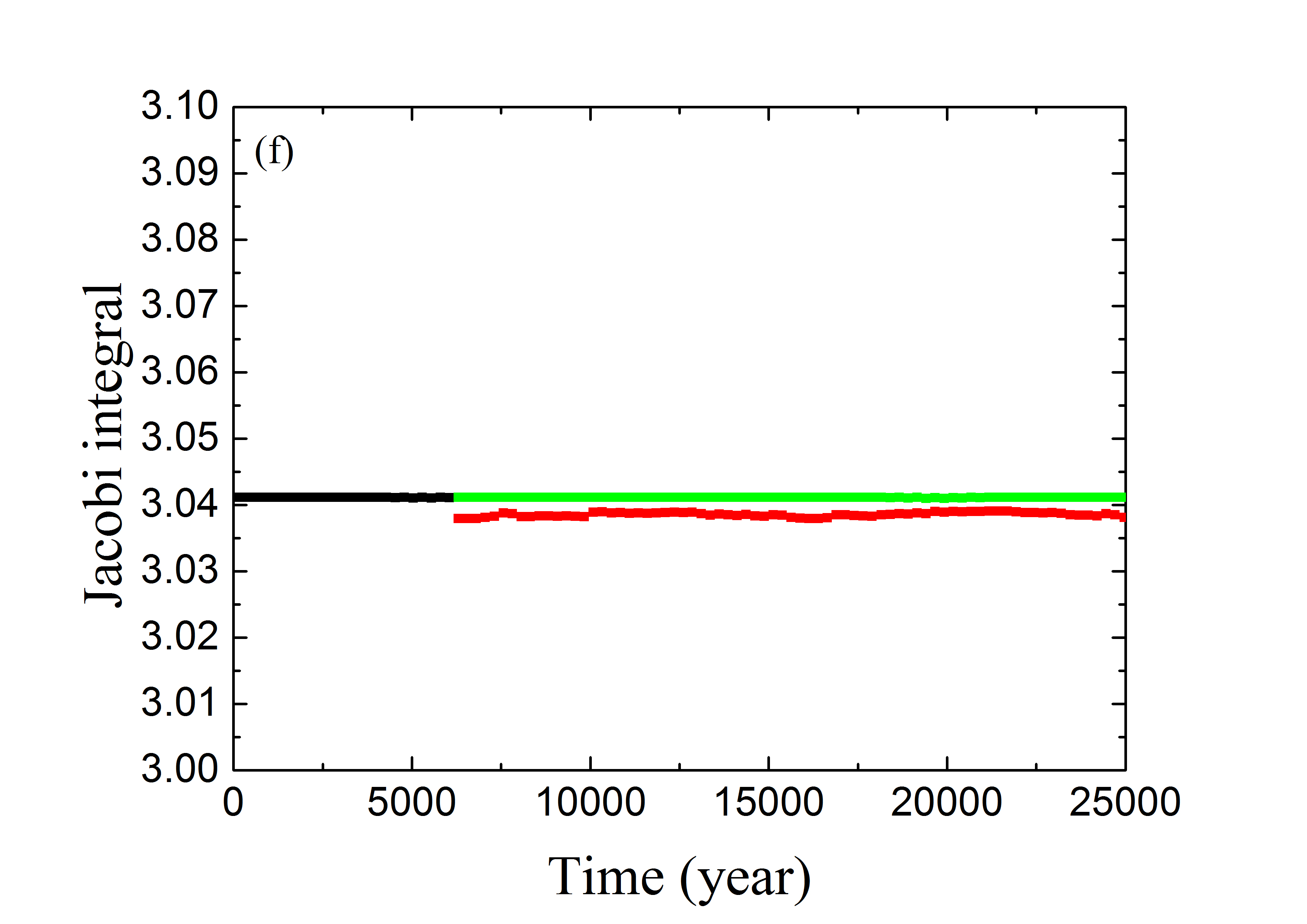}
  \end{minipage}
    \caption{For the case of $e=0.1$, three examples of the temporal evolution of the resonant angles $\sigma$ (left column) and Jacobi integral (right column) for particles from the validation set. The initial values for the first 6250 yrs are provided by the numerical integration (black dots). For the following 18750 yrs, the orbital parameters are predicted by the trained ANN using \textit{Train} \Rmnum3 (red dots) and calculated by extending numerical integration (green dots), respectively. (upper panels) The particle tp1 starting with $\sigma_0=79^\circ$ achieves the minimum loss of 0.007; (middle panels) The particle tp2 with $\sigma_0=126^\circ$ has a loss of 0.035; (bottom panels) The particle tp3 with $\sigma_0=60^\circ$ has a loss of 0.044. In the left panels, the best-fit sinusoid to predicted $\sigma$ values (i.e. the discrete red dots) is plotted as the yellow curve.}
  \label{ResArg}
\end{figure*}

For training the ANN by using either \textit{Train} \Rmnum2 or \textit{Train} \Rmnum3, the epoch number is also set to be 3000 as the one used for \textit{Train} \Rmnum1. Every once in a while, we visually checked the profile of loss validation $\mathcal{L}_{valid}$ in order to see how the ANN performs. At the end of both the different training scenarios, the overall $\mathcal{L}_{valid}$ values have already ceased to decrease. The individual loss curves are plotted in Fig. \ref{loss}.

The loss profile corresponding to \textit{Train} \Rmnum2 is presented by the green curve in Fig. \ref{loss}. We can see that the losses $\mathcal{L}_{valid}$ at both ends are about 0.05, which is $\sim17$\% smaller than \textit{Train} \Rmnum1 (blue curve); while the central losses increase a little bit. This trend is natural, because of the different number distributions of \textit{Train} \Rmnum1 and \textit{Train} \Rmnum2. The latter has a steeper number distribution ($\Sigma \propto (A_{\sigma})^{1}$) than the former ($\Sigma \propto (A_{\sigma})^{0}$) and its data is more adequate at larger $A_{\sigma}$ (equivalent to the initial resonant angle $\sigma_0$). Thereby we can justify the obvious improvement in training the ANN by adopting the set \textit{Train} \Rmnum2.

When we continue to sharpen the number distribution of the training samples to $\Sigma \propto (A_{\sigma})^{2}$, i.e. using the training set \textit{Train} \Rmnum3, the loss profile (red curve in Fig. \ref{loss}) still follows the trend described previously: a decrease on both ends but an increase in the central region. In this scenario we can see that the overall loss curve becomes rather flattened as the maximum $\mathcal{L}_{valid}$ is consistently at the level of $\sim 0.04$. At this point, we do not expect that the overall loss will be significantly reduced by an even steeper $\Sigma$.

One may notice the zig-zag shape of individual loss curves in Fig. \ref{loss}, specifically, there are peaks (minima) at approximately $\pm45^\circ$ ($\pm90^\circ$) away from the resonant center. Indeed, the profile of loss curve depends sensitively on the distribution of training data. The first layer of our ANN contains a group of $1\times1$ convolution filters which work essentially according to the local data distribution. The zig-zag shape means that the required local data distribution should be a bit different from those given in Table \ref{trainset}.  More precisely, the number densities of training samples in the vicinities of the peaks should be higher. There is another feature in Fig. \ref{loss}, as we can see the peaks get broader with steeper number distribution $\Sigma$ of training samples. Considering the region of $\sigma_0=90^\circ$-$180^\circ$, due to larger $\Sigma$, \textit{Train} \Rmnum3 has less training samples than \textit{Train} \Rmnum1 and \textit{Train} \Rmnum2. Thus insufficient samples lead to higher losses in this region, as shown by the broader peak of the red curve. Nevertheless, since the overall loss is acceptable, we would not further fine-tune the local distribution of training data.

\subsubsection{Prediction of resonant angle and resonant amplitude}

In order to display the performance of the ANN trained by the set \textit{Train} \Rmnum3, we present in Fig.\ref{ResArg} the orbital prediction for three representative examples from the validation set. The left column lists the time evolution of the resonant angle $\sigma$, which can validate the ANN's prediction effects of the orbital angles $\varpi$ and $\lambda$. As for the particle tp1 starting with the resonant angle $\sigma_0\approx79^\circ$, it achieves the minimum loss of $\mathcal{L}_{valid}\sim0.007$ (see the red curve in Fig. \ref{loss}). The evolution of this sample's $\sigma$ is presented in Fig. \ref{ResArg}a. The black dots represent the $\sigma$ values in the first $\frac{1}{4}\cdot T_{tot}=6250$ yrs as the initial conditions of the machine learning method. For the next $\frac{3}{4}\cdot T_{tot}=18750$ yrs, the evolving $\sigma$ are predicted by the trained ANN (red dots), and calculated by the numerical integration (green dots). The two solutions seem to match each other perfectly. However, the performance of the trained ANN may be relatively poor for the particles tp2 with $\sigma_0=126^\circ$ and tp3 with $\sigma_0=60^\circ$, as they have the largest losses of $\mathcal{L}_{valid}\sim0.04$ (see Fig. \ref{loss}). Figs. \ref{ResArg}b and \ref{ResArg}c show the evolution of the resonant angles of tp2 and tp3, respectively. It can be seen that the $\sigma$ values obtained from numerical integration (green) and machine learning prediction (red) could be modestly different from time to time. This issue is due to the current overall losses which can not be reduced anymore in the designed ANN. Some considerations in mind for the structure of the ANN may be helpful to further improve the machine learning's ability to predict the resonant trajectories. We will come back to this in the discussion of this paper.

\begin{figure}
 \hspace{0cm}
  \centering
  \includegraphics[width=9cm]{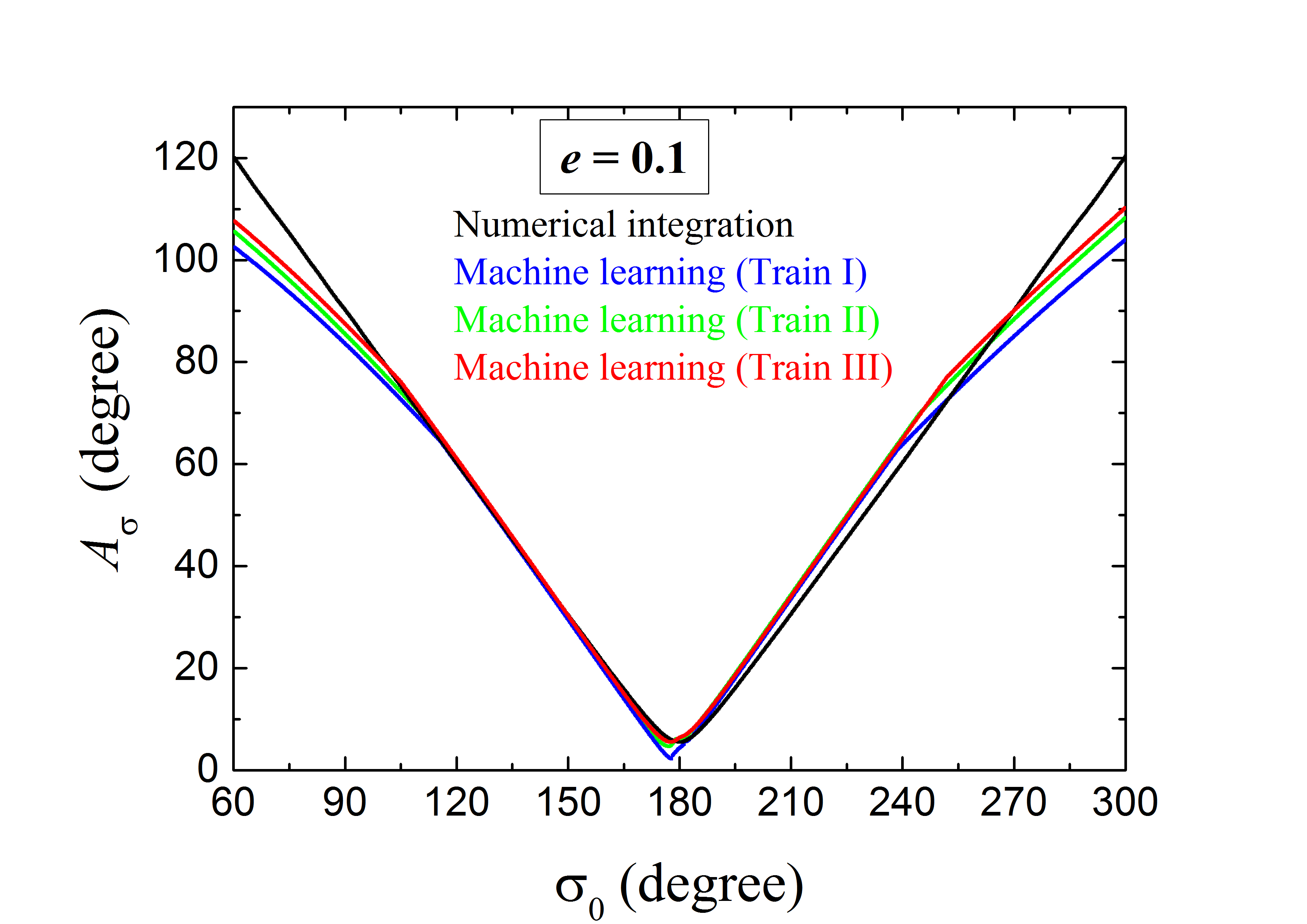}
  \caption{Resonant amplitudes $A_{\sigma}$ of particles from the validation set for $e=0.1$. The black curve indicates the numerical integration calculation, the coloured curves indicate the predictions of the ANNs trained by the sets \textit{Train} \Rmnum1 (blue),  \textit{Train} \Rmnum2 (green) and \textit{Train} \Rmnum3 (red). The results show that the machine learning associated to \textit{Train} \Rmnum3 works best, and it can effectively identify the particles as stable 2:3 resonators by the criterion $A_{\sigma}<120^\circ$.}
  \label{ample01}
\end{figure}

Nevertheless, for the sake of identifying the resonant status, we merely need to measure the maximum deviation of $\sigma$ from the resonant centre, i.e. the resonant amplitude $A_{\sigma}$. Then, a particle can be regarded as being stable in the 2:3 MMR if $A_{\sigma}<120^\circ$, and such a criterion may be less dependent on the error of $\sigma$ at a specific time. As shown in Figs. \ref{ResArg}b (for tp2) and  \ref{ResArg}c (for tp3), although there is an obvious phase difference between the green (from numerical integration) and red curves (from machine learning), the amplitudes of these two sine-like curves (equivalent to $A_{\sigma}$) could be comparable. Then for each of all the 1000 samples from the validation set, we determine its $A_{\sigma}$ value from a total of 25000 yrs of evolution. The choice of this timescale is reasonable since it is longer than the typical period of a full $\sigma$-libration cycle. In order to better measure the libration amplitude $A_{\sigma}$, instead of using the maximally-deviant point, we carried out the  sinusoidal fits to the discrete (red) points. In Figs. \ref{ResArg}a-\ref{ResArg}c, the best-fit sinusoid is plotted as the yellow curve, and the amplitude $A_{\sigma}$ is derived accordingly from the fitting procedure.

Fig. \ref{ample01} shows the resonant amplitudes $A_{\sigma}$ of the validation particles, acquired by numerical integration (black), and the ANNs trained by the training sets \textit{Train} \Rmnum1 (blue), \textit{Train} \Rmnum2 (green) and \textit{Train} \Rmnum3 (red). It is interesting to note that, for the orbits with $\sigma_0\in (110^\circ,250^\circ)$ in the central region around the resonant centre at $180^\circ$, there is a near perfect match between the predicted and real $A_{\sigma}$. We further notice that the machine learning curves are slightly above the numerical method curve for $\sigma_0=180^\circ$-$250^\circ$, while they seem to match better for the other side of  $\sigma_0=110^\circ$-$180^\circ$. It seems that the machine has not learned the symmetrical characteristic of the 2:3 MMR with respect to the resonant center at $180^\circ$, as displayed in Fig. \ref{pendu}. A likely cause is that the $1\times1$ convolutions embedded in our ANN focus on each element in the dataset, and they are not exactly symmetric. Nevertheless, this asymmetry can be neglected when the designed ANN is applied to identify the resonant objects.

Besides, we can see in Fig. \ref{ample01} that, the accuracy of prediction becomes lower at either edge of this stable resonance zone. In the regions far away from the resonant center, i.e. of $\sigma_0<110^\circ$ and $>250^\circ$, the relative differences between the machine learning and numerical integration curves are $\lesssim10$\%. At this point, the trained ANN could also supply comparable $A_{\sigma}$ and somewhat smaller than $120^\circ$. Consequently, the associated particles will still be classified as the stable 2:3 resonators. Through a closer look at the $A_{\sigma}$ profiles, the case of \textit{Train} \Rmnum1 performs relatively poor, while the case of \textit{Train} \Rmnum3 seems to work best, as evaluated by the performance of the ANN at largest values of $A_{\sigma}$. This result is consistent with our previous discussion about the loss profiles for these three cases in Section 3.3. Hereafter, the ANN constructed from \textit{Train} \Rmnum3 will be referred as ``best-trained ANN'', and applied to the resonant objects with various orbital elements.

\begin{figure}
 \hspace{-0.5cm}
  \centering
  \includegraphics[width=10cm]{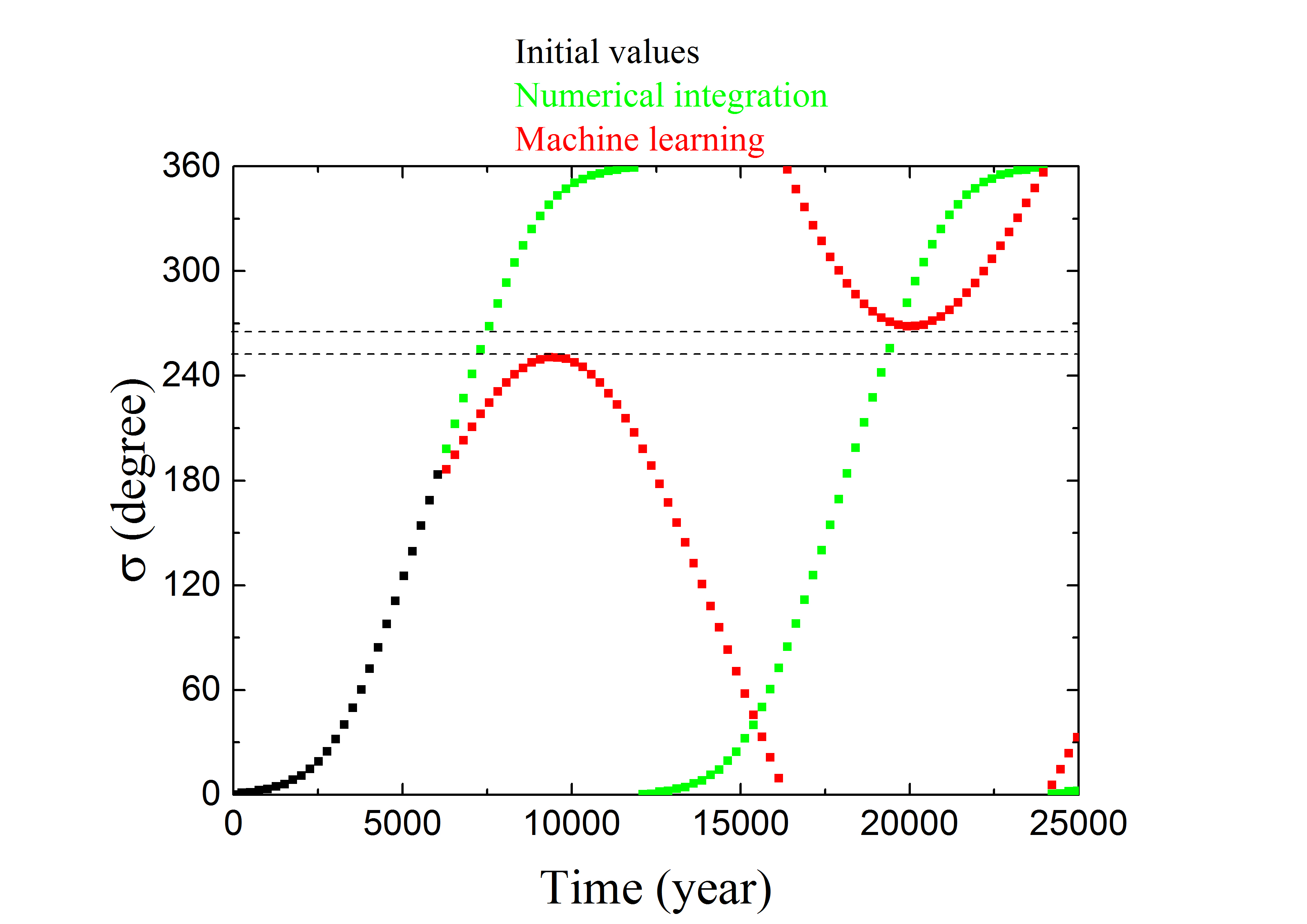}
  \caption{Same as Fig. \ref{ResArg} (left column panels), but for a non-resonant particle. This particle goes through the best-trained ANN associated to \textit{Train} \Rmnum3. It can be seen that the predicted resonant angle $\sigma$ (red dots) can reach the value of 0, which is $180^\circ$ away from the resonant centre, indicating non-resonant motion. For reference, the horizontal dashed lines are plotted at $\sigma=250^\circ$ and $268^\circ$ respectively.}
  \label{nonres}
\end{figure}

To additionally evaluate the performance of the best-trained ANN, the right column of Fig. \ref{ResArg} lists the time evolution of the Jacobi integral $C$ for the validation samples tp1, tp2 and tp3. The colours have the same meanings as in the left column. In terms of $a$ and $e$, the value of $C$ can be expressed as
\begin{equation}
C(a, e)=\frac{1}{a}+2\sqrt{a(1-e^2)}+\mathcal{O}(\mu).
\label{jacobi2}
\end{equation}
We note that this approximate $C(a, e)$ is different from the exact $C(x, y, \dot{x}, \dot{y})$ in equation (\ref{jacobi}) only by the order of $\mu\sim10^{-5}$. Then the values of $C(a, e)$ can be used to validate the prediction of the other two orbital elements $a$ and $e$. We find that, for any of the three resonant examples, the variation of the predicted $C$ (red dots) at different time-points is extremely small, of the order of $<$ 0.2\%. Although the predicted $C$ could be slightly apart from the labels (green dots), the conservation of the Jacobi integral does hold well.

\subsubsection{Generalization on the non-resonant case}

Having considered the resonant objects, we now turn to examining what a non-resonant object looks like when it goes through our best-trained ANN. We generated some non-resonant objects in the validation set, by giving them $\sigma_0\in(179^\circ,180^\circ)$, while keeping the other orbital parameters the same. In the numerical integration, these objects are classified to be non-resonant as they have their resonant angles $\sigma$ circulating during the evolution. An example is provided in Fig. \ref{nonres}. Given the initial values (black), we show its subsequent $\sigma$ evolution from the numerical integration (green) and the machine learning prediction (red). It can be seen that the predicted $\sigma$ does reach the value of $0^\circ$, which is $180^\circ$ away from the resonant centre, indicating a resonant amplitude of $A_{\sigma}=180^\circ$ according to the usual definition. Thus our best-trained ANN is also capable of effectively classifying the non-resonant population.

One may notice in Fig. \ref{nonres} that, the predicted $\sigma$ (red curve) is not exactly circulating as there is a narrow gap between $\sigma=250^\circ$ and $268^\circ$, indicated by the dashed lines. The local peak of $\sigma=250^\circ$ appears at the time about 10000 yrs, which is close to half the libration period of the 2:3 resonance. At this moment, the evolution of $\sigma$ turns to decreasing, in contrast to 
the realistic (circulating) behaviour represented by the green curve which is increasing.  Such a profile switch is due to the fact that only the resonant trajectories were used to train the ANN, and they tell the non-resonant ones to behave alike, i.e. the monotonic increase of $\sigma$ should not last longer than half of the 2:3 resonant period (see Fig. \ref{ResArg}). Nevertheless, as said above, the usual criterion identifying the non-resonators with $A_{\sigma}=180^\circ$ would not be affected at all.

\subsection{The cases of larger $e=0.2$ and $e=0.3$}

\begin{figure}
 \hspace{0cm}
  \centering
  \includegraphics[width=9cm]{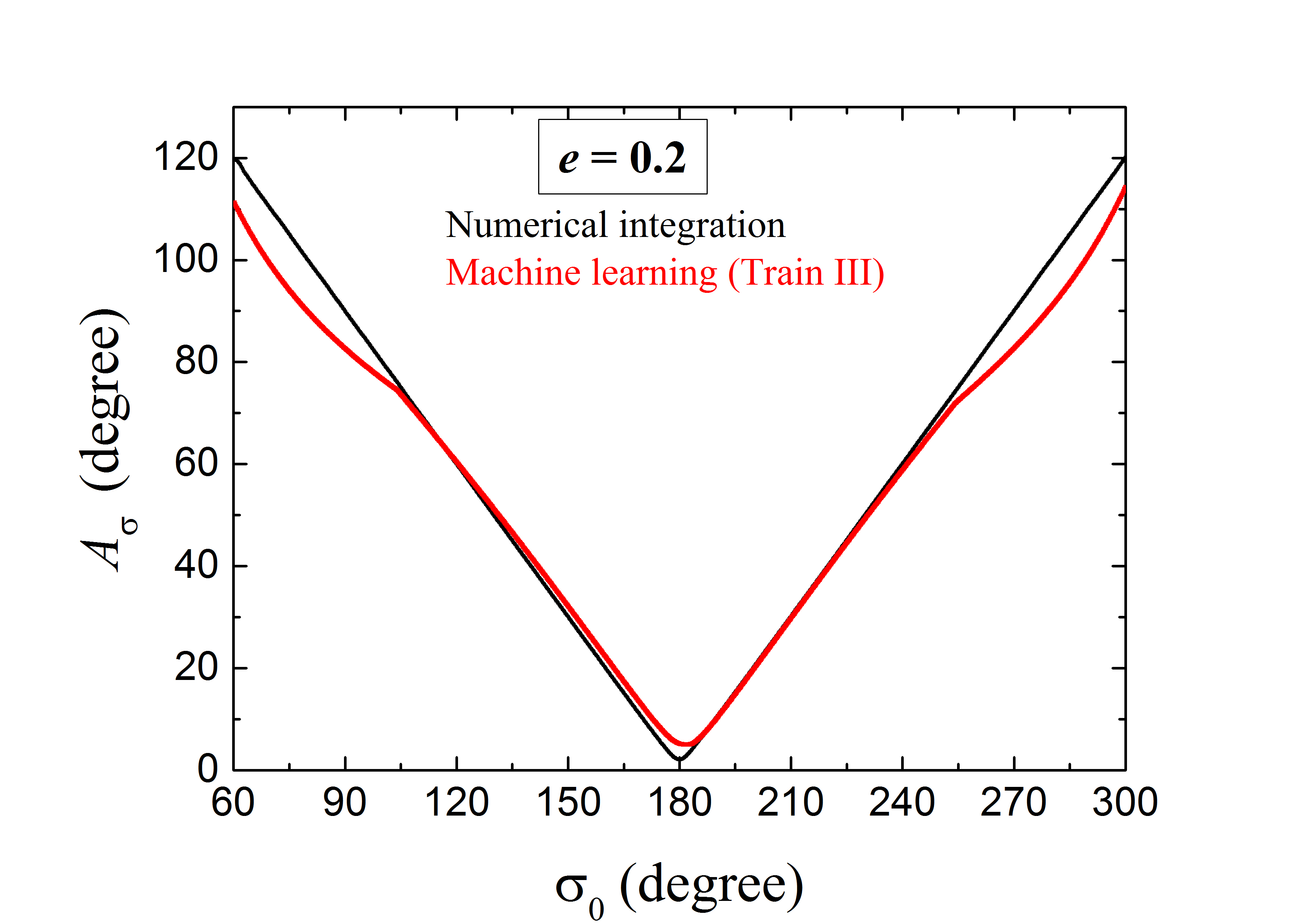}
  \caption{Same as Fig. \ref{ample01}, but for $e=0.2$. Note that only the prediction of the best-trained ANN associated to \textit{Train} \Rmnum3 is presented.}
  \label{ample02}
\end{figure}

\begin{figure}
 \hspace{0cm}
  \centering
  \includegraphics[width=9cm]{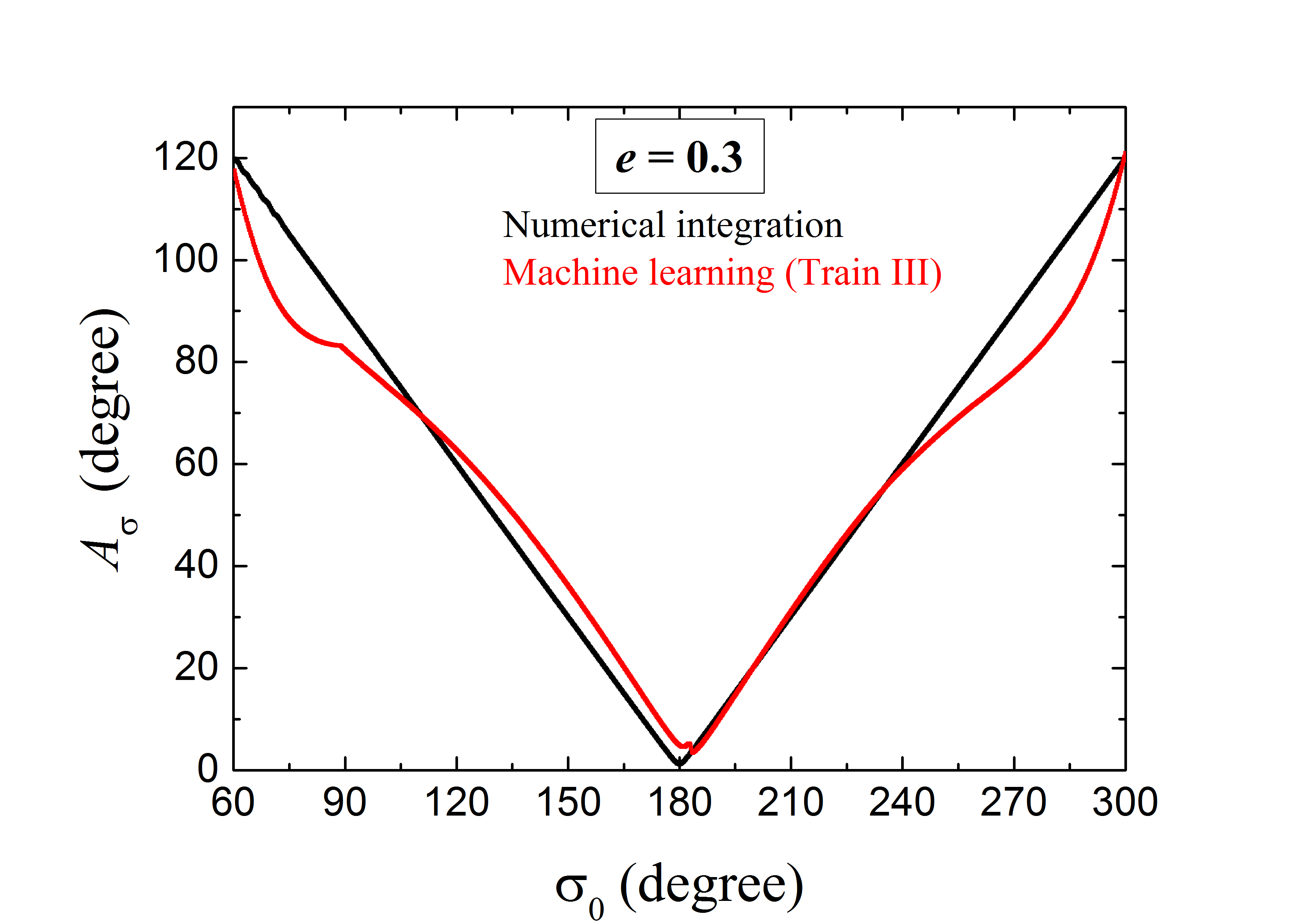}
  \caption{Same as Fig. \ref{ample02}, but for $e=0.3$.}
  \label{ample03}
\end{figure}

As shown above, our best-trained ANN can well predict the resonant status for trajectories with small eccentricity $e=0.1$. It is also of great significance to investigate whether this ANN model could be applied to more eccentric cases. Until now, a large fraction of the 2:3 resonant KBOs have been observed on trajectories with large $e$, as high as about 0.3.

For the training and validation samples, the initial set of parameters $(a, \varpi, \lambda)$ has the same values as the one in the $e=0.1$ case, but their eccentricities are set to the other two representative values of $e=0.2$ and $e=0.3$. Then we re-trained the  best ANN model, which involves a number density of the training samples as $\Sigma \propto (A_{\sigma})^{2}$ (i.e. \textit{Train} \Rmnum3, see Table \ref{trainset}). During the training process, we found that the training losses still decrease at a similar rate as before. After 3000 training epochs are processed, the training losses can hardly decrease and the validation losses don't seem to increase. Therefore, we also evaluate the performance of the obtained ANN on the validation set at epoch 3000.

Fig. \ref{ample02} and Fig. \ref{ample03} present the results for the validation particles with $e=0.2$ and $e=0.3$, respectively. We observe that the overall profiles of the resonant amplitudes $A_{\sigma}$ predicted by the ANN (red curve) and calculated from the numerical integration  (black curve) seem alike. By comparing with the case of $e=0.1$ (see the red curve in Fig. \ref{ample01}), one would immediately notice that: (1) all the stable 2:3 resonators with $A_{\sigma}<120^\circ$ can be effectively identified by the machine learning method. (2) Similarly, the machine learning prediction of $A_{\sigma}$ can be quite accurate for resonant orbits having small to medium $A_{\sigma}$, i.e. with $\sigma_0\in(110^\circ,250^\circ)$. (3) At large $A_{\sigma}$, modest but comparable relative difference between the predicted and numerical curves could arises, at the level of $\lesssim$ 10\%-15\%. As we have said before, this error is due to less information for the particles farther away from the resonant centre.

However, in the resonant region at large $A_{\sigma}$, the profiles of the predicted $A_{\sigma}$ are somewhat dissimilar for different $e$. Regarding the symmetry of the $A_{\sigma}$ curve, we just consider the validation particles with $\sigma_0<110^\circ$, i.e. the ones located at the very left part in Figs. \ref{ample01}, \ref{ample02} and \ref{ample03}. For the sake of clarity, the discrepancy of $A_{\sigma}$ between the machine learning prediction and the numerical integration calculation at a specific $\sigma_0$, will be denoted by $\Delta A_{\sigma}$. In the cases of $e=0.1$, $e=0.2$ and $e=0.3$, the maximum $\Delta A_{\sigma}$ are $12.4^\circ$, $11.1^\circ$ and $16.8^\circ$, respectively. It seems that the worst prediction comes from the most eccentric orbits. On the contrary, when we look at the left end of the $A_{\sigma}$ curve (i.e. $\sigma_0=60^\circ$), the validation particle with $e=0.3$ has the most accurate prediction of $\Delta A_{\sigma}=2.3^\circ$. In total, if considering the overall region of $\sigma_0=60^\circ$-$110^\circ$, the mean values of relative differences of $A_{\sigma}$ are roughly the same of $\sim$5\%-9\% for $e=$ 0.1-0.3. Accordingly, for predicting the libration behaviours of particles with different $e$, we deem that the general abilities of the trained ANNs could be similar. As a matter of fact, we checked the total losses on the entire validation sets in the cases of $e=0.1$, $e=0.2$ and $e=0.3$, and found that their values are of the same order of magnitude.

In summary, our results imply that the machine learning method could effectively identify the dynamical status of the particles as stable resonant or non-resonant, at a computational expense much lower than the one of the numerical integration. The application of a trained ANN may help screen a huge number of KBOs expected to be discovered in the near future and select a small fraction of them to be resonant candidates, which are then further confirmed by numerical integrations. This procedure could considerably reduce the computational time for the classification of the resonant KBOs.


\section{Conclusions}


Some very recent works tried to use machine learning to solve integrable Hamiltonian systems \citep{Grey2019} and the general 3-body problem \citep{Breen2020}. The former case essentially contains only regular motion and the trained ANN was claimed to be capable of predicting accurate dynamics over a long timespan, while the latter case involves chaotic motion, thus the accuracy of the ANN's prediction is strongly dependent on the system's evolution time, which should be quite limited. In this paper, we investigate the performance of machine learning predictions for the 2:3 MMR behaviour in the Sun+Neptune+particle PCR3BP. For small eccentricity $e=0.1$, the 2:3 resonator exhibits regular motion in the non-integrable system, which can be regarded as an intermediate case. Considering the stability of the realistic 2:3 resonant KBOs, we only take into account the samples having resonant amplitudes $A_{\sigma}<120^\circ$ \citep{Li2014}.

Providing the trajectory data over a time interval of $T_{tot}=25000$ yrs by numerical integration, we trained the ANN to learn the behaviour of the particles in the 2:3 MMR with Neptune. In supervised learning, we managed to design an appropriate pattern of the training set (i.e. \textit{Train} \Rmnum3 in Table \ref{trainset}) and an ANN containing convolutional and fully connected layers. Our results show that by using the initial data in the first $\frac{1}{4}\cdot T_{tot}=6250$ yrs, the best-trained ANN can predict the trajectories of the 2:3 resonators for the subsequent $\frac{3}{4}\cdot T_{tot}=18750$ yrs. As for the evolving resonant angle $\sigma$, the machine learning prediction and numerical integration calculation seem to match each other nicely and the relative errors could be as small as only of a few degrees. In addition, the predicted Jacobi integral, as a function of the orbital semimajor axis and eccentricity, could also be well conserved.

Besides the orbit prediction, we aim to identify the stable 2:3 resonators by means of measuring the resonant amplitude $A_{\sigma}$. Combining the given 6250 yr data and the predicted 18750 yr data, we have the evolving $\sigma$ over a total timescale of 25000 yrs, which is longer than a full libration cycle of the 2:3 MMR. We find that the said best-trained ANN could effectively classify the resonant population with $A_{\sigma}<120^\circ$ in the validation set at a very confidential level. Subsequently, our ANN model is applied to the 2:3 resonators on more eccentric orbits with $e$ up to 0.3, and the machine learning method still works very well in predicting their resonant amplitudes $A_{\sigma}$.

\begin{figure}
 \hspace{0cm}
  \centering
  \includegraphics[width=9cm]{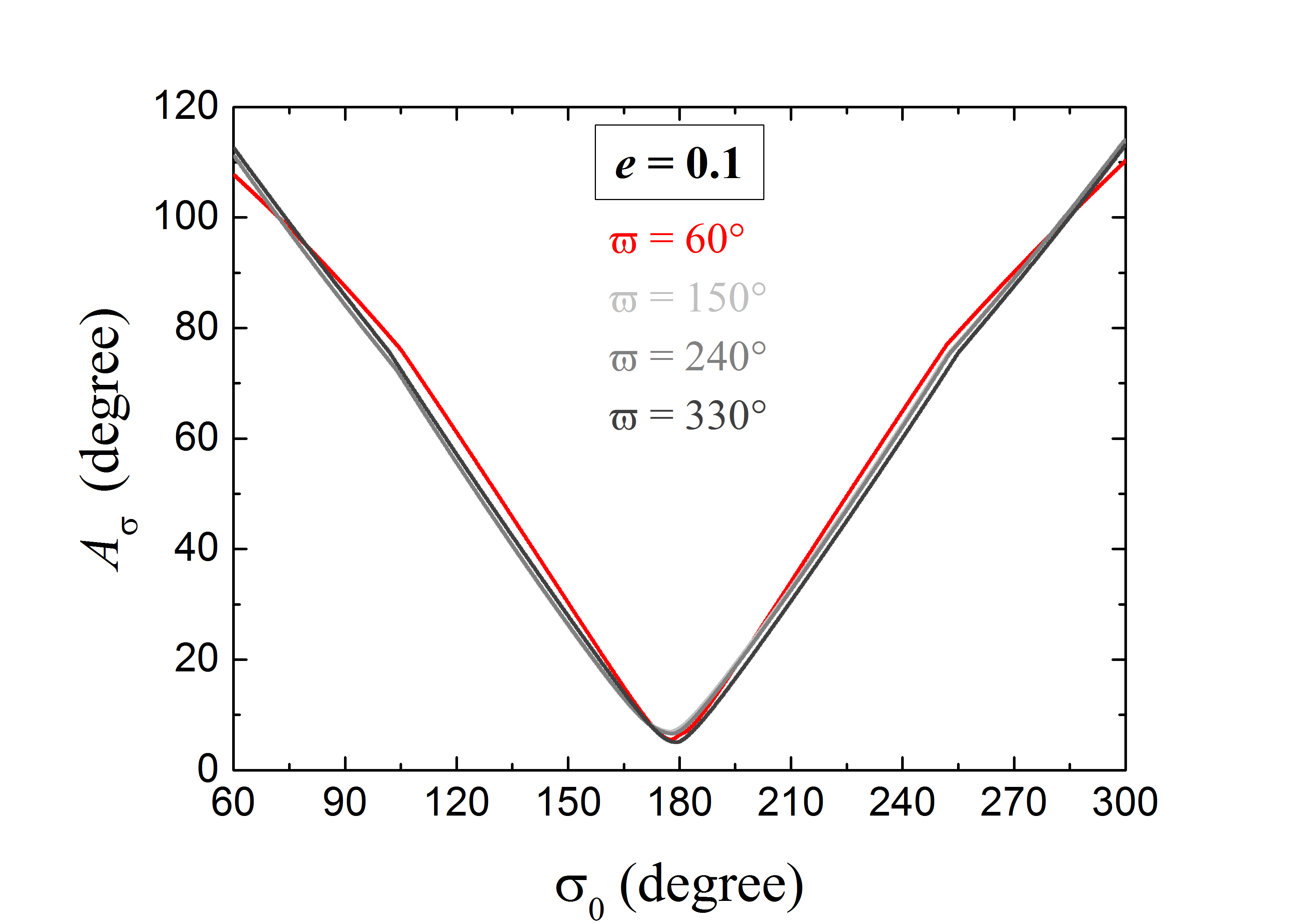}
  \caption{Resonant amplitudes $A_{\sigma}$ predicted by the trained ANN associated to \textit{Train} \Rmnum3. Each curve corresponds to a different longitude of perihelion value $\varpi$. The red curve corresponding to $\varpi=60^\circ$ is extracted from Fig. \ref{ample01}.  For all cases, $e=0.1$.}
  \label{v3dperi}
\end{figure}

In conclusion, we propose that the machine learning method may serve as a fast approach in identifying the dynamical status of the objects in the stable 2:3 MMR with Neptune. This way, we can save as much as $75\%$ of computational time compared to making use of numerical integrations. With this advantage, machine learning may quickly produce a preliminary set of candidate resonators for the dynamical classification of the realistic KBOs. Definitely, much more work is needed to be done to develop the ANN more practically. 

\section{Discussion}

There is a concern about the construction of our best-trained ANN, i.e. the choice of the longitude of perihelion $\varpi$. As described in equation (\ref{ResAng}), once $\varpi$ is fixed, the resonant angle $\sigma$ is equivalent to sampling the mean longitude $\lambda$ of the particle, because Neptune's mean longitude $\lambda_N$ is initially equal to 0. May an expanded $\varpi$ sampling have consequences on our main results? In order to explicitly evaluate the associated impact on the machine learning prediction for $A_{\sigma}$, we here perform additional runs starting with distinct $\varpi$. Having already considered $\varpi=60^\circ$, we choose three additional values for $\varpi$: $\varpi=150^\circ$, $240^\circ$ and $330^\circ$. These four values are representative since they are distributed evenly in the entire region of $0-360^\circ$. In the case of $e=0.1$, we re-trained the best ANN associated to \textit{Train} \Rmnum3. These results are presented in Fig. \ref{v3dperi}. The red curve in Fig. \ref{v3dperi} is extracted from Fig. \ref{ample01}, referring to the previous ANN-estimated $A_{\sigma}$ for $\varpi=60^\circ$. The grey curves corresponds to the other three values of $\varpi$, and they are found to nearly overlap with the red one. Hence, the structure of our ANN model would not be affected by the choice of $\varpi$. This outcome is consistent with what we theoretically argued in Section 2.2, i.e. it is reasonable to choose an arbitrary value of $\varpi$ for constructing the ANN. As a result, as long as $\varpi$ is densely sampled in the range from 0 to $360^\circ$, we may train a more generic ANN. However, due to the computational limitations, currently it is not easy to fulfill this task.

We noticed in Fig. \ref{ResArg} that, although the predicted Jacobi integral is almost conserved, the small discrepancy with the numerical integration calculation could be a potential aspect of the ANN improvement. It should be noted that the PCR3BP is actually a Hamiltonian system and the Jacobi integral is sometimes called the integral of relative energy. This conservation could be used to improve the prediction accuracy, as \citet{Breen2020} did for the general 3-body problem. They added a projection layer to better preserve the Hamiltonian quality during the training of ANN. This layer adjusts the coordinates by minimizing a parameter related to the energy error. Similarly, the consideration of a conserved quantity may also be applied to study other dynamical systems by the ANNs. A new publication even suggests that, by using the trajectory data from unknown dynamical systems, the machine learning algorithm could automatically discover the conserved physical quantities \citep{Liu2021}.

In the framework of the PCR3BP with two degrees of freedom, the resonant motion has two fundamental frequencies. One frequency corresponds to the rate of change of the resonant angle, i.e. $\dot{\sigma}$. The other frequency referring to the secular evolution is associated to the rate of change of the longitude of perihelion, i.e. $\dot{\varpi}$. In the stage of data pre-treatment, as a component of the ANN, we may use the convolution kernel as a filter to find these two main frequencies at a rather high level of accuracy. Due to the perturbation from Neptune, the resonant motion still contains additional terms with much higher frequencies, denoted by $f_k$ $(k=1, 2, \dots)$. Now, if we require the overall accuracy to be around $80\%$, we only need to obtain the low frequencies $\dot{\sigma}$ and $\dot{\varpi}$ with an accuracy of $80\%$, which is achievable. But if we desire the total accuracy to reach $\sim90\%$, we have to let the ANN to efficiently learn the behaviour of the motion associated to the higher frequencies $f_k$. We suppose that the pre-treatment by means of extracting the fundamental frequencies may help us further reduce the loss on the validation set and increase the prediction accuracy. 

So far, we have investigated whether the ANN could learn the motion of the ``synthetic'' 2:3 resonant objects. Similarly to the usual approaches, we first considered the planar case, which requires lower computational capacity. Only if the machine learning method is found to be effective, it is possible to be further developed and generalised to realistic KBOs, e.g. for objects on inclined orbits. Accordingly, the orbital inclination and ascending node are to be included in the ANN model. In order to add two more dimensions in the training data, we need to train at least $10^6$ orbits to maintain the current prediction accuracy. In higher dimensional data space, overfitting is more likely to happen and the gradient descent optimisation may also encounter problems. A similar challenge arises in video classification, as one deals with high dimensional images. Adding several different fusion layers could be a strategy \citep{karpathy2014}.  Regularisation \citep{siri19}, or Microsoft's Automated Machine Learning on Azure\footnote{https://docs.icrosoft.com/en-us/azure/machine-learning/concept-manage-ml-pitfalls} could be additional possibilities. It is important to point out that, the structure of the ANN model strongly relies on the data distribution. Since 6-dimensional data is required to describe the inclined trajectories, the network structure could be quite different from that we obtained. Our next step is to find a suitable ANN, while it could cost substantial time to achieve this goal.


\section*{Acknowledgments}

This work was supported by the National Natural Science Foundation of China (Nos. 11973027, 11933001 and 11601159), and National Key R\&D Program of China (2019YFA0706601). We would also like to express our sincere thanks to the anonymous referee for the valuable comments.  

\section*{Data Availability}

The data underlying this article are available in the article and in its online supplementary material.

\end{document}